\documentstyle[aps,floats,eqsecnum,psfig]{revtex}

\begin{document}
\draft

\title{Resonant transducers for a spherical GW antenna}

\author{J. Alberto Lobo and M. Angeles Serrano}
\address{Departament de F\'\i sica Fonamental \\
         Universitat de Barcelona, Spain.}

\date{\today}
\date{3 June 1997) \\ (Revised \today}

\maketitle

\begin{abstract}

Apart from omnidirectional, a solid elastic sphere is a multimode and
multifrequency device for the detection of Gravitational Waves (GW).
Motion sensing in a spherical GW detector thus naturally requires a
multiple set of transducers attached to its surface at suitable
locations. If these transducers are of the resonant type then their
motion couples to that of the sphere, and the {\it joint dynamics\/}
of the system has to be properly understood before reliable conclusions
can be drawn from its readout. In this paper we address the problem of
the coupled motion of a solid elastic sphere and a set of resonators
attached to its surface in full theoretical rigor. A remarkably elegant
and powerful scheme is seen to emerge from the general equations which
shows with unprecedented precision how coupling takes place as a series
function of the small ``coupling constant'' $\eta\/$, the ratio of the
resonators' average mass to the sphere's mass. We reassess in the new
light the response of the highly symmetric truncated icosahedron layout
(the {\sl TIGA\/}), and also present a new proposal (the pentagonal
hexacontahedron, or {\sl PHC\/}), which has less symmetry but requires
only 5 rather than 6 transducers. We finally address the question of how
the system characteristics are affected by slight departures from perfect
spherical symmetry and identity of resonators, and find it to be quite
{\it robust\/} against such small failures. In particular, we recover with
fully satisfactory accuracy the reported experimental frequencies of the
reduced scale prototype detector constructed and tested at {\sl LSU\/}.

\end{abstract}


\section{Introduction}

The idea of using a solid elastic sphere as a gravitational wave (GW)
antenna is almost as old as that of using cylindrical bars: as far back
as 1971 Forward published a paper \cite{fo71} in which he assessed some
of the potentialities offered by a spherical solid for that purpose. It
was however Weber's ongoing philosophy and practice of using bars which
eventually prevailed and developed up to the present date, with the highly
sophisticated and sensitive ultracryogenic systems currently in operation
---see \cite{amaldi} and \cite{gr14} for rather detailed reviews and
bibliography. With few exceptions \cite{ad75,wp77}, spherical detectors
fell into oblivion for years, but interest in them strongly re-emerged in
the early 1990's, and an important number of research articles have been
published since which address a wide variety of problems in GW spherical
detector science. At the same time, international collaboration has
intensified, and prospects for the actual construction of large spherical
GW observatories (in the range of $\sim$100 tons) are being currently
considered in several countries \cite{coll}, even in a variant {\it hollow\/}
shape \cite{vega}.

A spherical antenna is obviously omnidirectional but, most important, it
is also a natural {\it multimode\/} device, i.e., when suitably monitored,
it can generate information on the GW amplitudes and incidence direction
\cite{wp77,nadja}, a capability which possesses no other individual GW
detector, whether resonant or interferometric. Furthermore, a spherical
antenna could also reveal the eventual existence of {\it monopole\/}
gravitational radiation, or set thresholds on it \cite{maura}. The
theoretical explanation of these facts is to be found in the unique
matching between the GW amplitude structure and that of the sphere
oscillation eigenmodes: a general {\it metric\/} GW generates a
{\it tidal\/} field of forces in an elastic body which is given in terms
of the ``electric'' components $R_{0i0j}(t)$ of the Riemann tensor at its
centre of mass by the following formula \cite{lobo}:

\begin{equation}
   {\bf f}_{\rm GW}({\bf x},t)\ \ \ =
    \sum_{\stackrel{\scriptstyle l=0\ {\rm and}\ 2}{m=-l,...,l}}\,
    {\bf f}^{(lm)}({\bf x})\,g^{(lm)}(t)    \label{1.1}
\end{equation}

where ${\bf f}^{(lm)}({\bf x})$ are pure ``tidal form factors'', while
$g^{(lm)}(t)$ are suitable linear combinations of the Riemann tensor
components $R_{0i0j}(t)$ which carry all the {\it dynamical\/} information
on the GW's monopole ($l\/$\,=\,0) and quadrupole ($l\/$\,=\,2) amplitudes.

On the other hand, a free elastic sphere has two families of oscillation
eigenmodes, so called {\it toroidal\/} and {\it spheroidal\/} modes, and
modes within either family group into ascending series of $l\/$-pole
harmonics, each of whose frequencies is (2$l\/$+1)-fold degenerate ---see
\cite{lobo} for full details. It so happens that {\it only\/} monopole
and/or quadrupole spheroidal modes can possibly be excited by an incoming
{\it metric\/} GW \cite{bian}, and their GW driven amplitudes are directly
proportional to the wave amplitudes $g^{(lm)}(t)$ of equation (\ref{1.1}).
It is this very fact which makes of the spherical detector such a natural
one for GW observations \cite{lobo}. In addition, a spherical antenna has
a significantly higher absorption {\it cross section\/} than a cylinder of
like fundamental frequency, and also presents good sensitivity at the
{\it second\/} quadrupole harmonic \cite{clo}.

But GW excitations are {\it extremely weak\/} \cite{schu}, and so a suitable
readout system must be {\it added\/} to the sphere in order to monitor its
motions and quantitatively assess their magnitude and physical significance.
In cylindrical bars, current state of the art technology is based upon
{\it resonant transducers\/} \cite{rapa,jpr,paik,pia,naut}. A resonant
transducer consists in a small (compared to the cylinder) mechanical device
possessing a resonance frequency accurately tuned to that of the cylinder.
This {\it frequency matching\/} causes back-and-forth {\it resonant energy
transfer\/} between the two bodies (bar and resonator), which results in
turn in {\it amplified\/} oscillations of the smaller resonator. The
{\it energy\/} amplification factor is $M_{\rm resonator}$/$M_{\rm bar}$,
hence the {\it amplitude\/} amplification is the {\it square root\/} of this
mass ratio. Pre-electronics mechanical amplification is highly desirable,
given the exceedingly small magnitude of any expected GW signals arriving
in the observatory.

The philosophy of using resonators for motion sensing is directly
transplantable to a spherical detector, but a {\it multiple\/} set rather
than a single resonator is required if its potential capabilities as a
multimode system are to be exploited to satisfaction. The practical
feasibility of a multiple transducer readout system has been recently
demonstrated {\it experimentally\/} with encouraging success by
S.\ Merkowitz and W.\ Johnson at {\sl LSU\/}, where a 740 kg prototype,
milled in the shape of a {\it truncated icosahedron\/}, and endowed with
a highly symmetric set of 6 resonators, was put to test
\cite{jm93,jm95,phd,jm97}. Their authors call this system {\sl TIGA\/},
an acronym for {\sl T\/}runcated {\sl I\/}cosahedron {\sl G\/}ravitational
{\sl A}ntenna.

One of the pillars of the success of this prototype experiment has been
M\&J's ability to give an adequate {\it theoretical interpretation\/} of
their experimental results. M\&J's model of the coupled dynamics of sphere
and resonators is based upon the hypothesis that {\it only\/} quadrupole
excitations of the sphere's modes need to be considered in a GW detector,
{\it even if\/} resonators are attached to the sphere's surface. This
extrapolation of a result which, as we have just mentioned, does hold
exactly for a {\it free\/} sphere, produces remarkably accurate predictions
of the {\it coupled\/} system behaviour, too, even though interactions of the
resonator set with the other, non-quadrupole sphere's modes are neglected
from the beginning. This is a strong indication that such neglected effects
are {\it second order\/} for the accuracy of the experimental data.

But what does ``second order'' precisely mean? The answer to such question
requires the construction of a more elaborate model, which should be
suitable to address in a systematic way any dynamical effects, and to
{\it quantitatively\/} assess their real importance. The interest of a
more sophisticated analysis is to understand and make clear the nature of
a given approximation scheme, as well as to enable further refinement of
it, if eventually required; its practical relevance is related to the
reasonable expectation that future real spherical GW detectors will make
use of extremely precise measurement techniques, likely to be rather
demanding as regards accurate theoretical modeling of the system.

The purpose and motivation of this paper is to present and develop such
a more refined mathematical model. We address the the problem of the
joint dynamics of a spherical elastic solid endowed with a set of
radial resonators, with as few as possible unwarranted hypotheses, and
with the objective to determine the system response to any interesting
signals, whether GWs or calibration inputs, with unlimited
mathematical precision. As we shall see, the model confirms and
generalises that of Merkowitz and Johnson \cite{jm93,jm97} in the sense
of making precise its actual range of applicability. More specifically,
we shall see that the solution to our general equations of motion can
be written as a {\it perturbative\/} series expansion in ascending
powers of the small {\it coupling constant\/} $\eta^{1/2}$
($\eta\/$\,$\equiv$\,$M_{\rm resonator}$/$M_{\rm sphere}$), whose
{\it lowest order\/} terms exactly correspond to Merkowitz and Johnson's
model. This is a key result, showing that the above alluded ``second
order'' effects are precisely order $\eta\/$ effects, or that M\&J's model
is accurate up to relative errors of this order.

Beyond this, though, a remarkably elegant, simple, and powerful algebraic
scheme will be seen to emerge from the theory, which neatly displays the
basic structure of the system for {\it completely general\/} resonator
distributions over the sphere's surface. Based on the resulting equations
we already advanced in references \cite{ls,lsc} a genuine transducer layout
with 5 rather than 6 resonators, which we propose to call {\sl PHC\/} (for
{\it pentagonal hexacontahedron\/}, the shape of the underlying polyhedron),
and which we shall also consider here in parallel with the more symmetric
{\sl TIGA\/} of Merkowitz and Johnson.

The paper will be structured as follows. Section 2 is devoted to present
the main hypotheses of the model and the general equations, whose reduction
to a tractable set is described in detail in section 3. Sections 4 and 5
contain a number of general, signal-independent results which apply to an
idealised system of perfect sphere and resonators; they constitute the
foundations for the analysis of more realistic instances, where one (or
more) of the idealised hypotheses partly fails. The system response to an
incoming GW is then addressed in section 6, where a complete study of
quadrupole {\it and\/} monopole radiation sensing is presented for
arbitrary transducer layouts. Still within the section, we apply our general
results to the important examples of {\sl TIGA\/} and {\sl PHC\/}. Section
7 addresses the problem of the system response to a simple
{\it calibration\/} signal, an impulsive hammer stroke, and in section 8 we
assess the consequences of system deffects relative to the idealised
perfection assumed that far; more specifically, we consider failures in
spherical symmetry and identity of resonator properties, and tolerances in
resonator location. In this section we also put to test our model's
predictions by confronting them with the reported experimental data obtained
in the {\sl TIGA\/} prototype experiment \cite{phd}, and see that agreement
between both (theory and experiment) is fully satisfactory in the given
theoretical and experimental conditions. The paper closes in section 9 with
a summary of conclusions.

\section{General equations}

\begin{figure}[htb]
\psfig{file=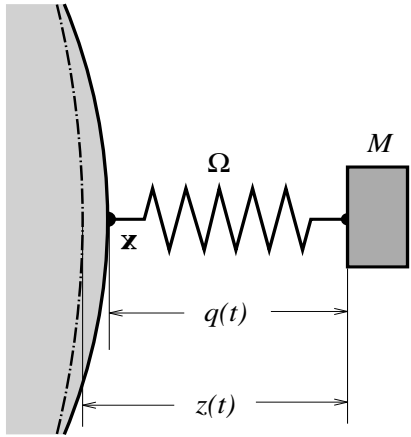,height=17cm,width=12cm,rheight=6.8cm,bbllx=-3cm,bblly=-7.4cm,bburx=14.2cm,bbury=19cm}
\caption{Schematic diagramme of the coupling model between a solid sphere
and a resonator. The notation is that in the text, but subindices have been
dropped for clarity. The dashed-dotted arc line on the left indicates the
position of the {\it undeformed\/} sphere's surface, and the solid arc its
{\it actual\/} position.  \label{fig1}}
\end{figure}

With minor improvements, we shall use the notation of references \cite{lobo}
and \cite{ls}, some of which is now briefly recalled. We consider a solid
sphere of mass $\cal M\/$, radius $R\/$, (uniform) density $\varrho\/$, and
elastic Lam\'e coefficients \cite{ll70} $\lambda$ and $\mu\/$, endowed with
a set of $J\/$ resonators of masses $M_a\/$ and resonance frequencies
$\Omega_a\/$ ($a\/$\,=\,1,\ldots,$J\/$), respectively. We shall model the
latter as {\it point masses\/} attached to one end of a linear spring,
whose other end is rigidly linked to the sphere at locations ${\bf x}_a\/$
---see Figure \ref{fig1}. The system degrees of freedom are given by the
{\it field\/} of elastic displacements ${\bf u}({\bf x},t)$ of the sphere
plus the {\it discrete\/} set of resonator spring deformations $z_a(t)$;
equations of motion need to be written down for them, of course, and this
is our next concern in this section.

We shall assume that the resonators only move radially, and also that
Classical Elasticity theory \cite{ll70} is sufficiently accurate for our
purposes\footnote{
We clearly do not expect relativistic motions in extremely small displacements
at typical frequencies in the range of 1 kHz.}.
In these circumstances we have \cite{ls}

\begin{mathletters}
\label{2.1}
\begin{eqnarray}
    \varrho \frac{\partial^2 {\bf u}}{\partial t^2} & = & \mu\nabla^2 {\bf u}
    + (\lambda+\mu)\,\nabla(\nabla{\bf\cdot}{\bf u}) + {\bf f}({\bf x},t)
    \label{2.1.a}  \\*[0.7 em]
    \ddot{z}_a(t) & = & -\Omega_a^2\,
    \left[z_a(t)-u_a(t)\right]+\xi_a(t)\ , \qquad a=1,\ldots,J
    \label{2.1.b}
\end{eqnarray}
\end{mathletters}

where ${\bf n}_a\/$\,$\equiv$\,${\bf x}_a/R\/$ is the outward pointing normal
at the the $a\/$-th resonator's attachement point, and

\begin{equation}
  u_a(t)\equiv{\bf n}_a\!\cdot\!{\bf u}({\bf x}_a,t)\ ,\qquad
  a=1,\ldots,J    \label{3.8}
\end{equation}

is the {\it radial\/} deformation of the sphere's surface at ${\bf x}_a\/$.
A dot (\,$\dot{}$\,) is an abbreviation for time derivative. The term in
square brackets in (\ref{2.1.b}) is thus the spring deformation ---$q(t)$ in
Figure \ref{fig1}.

${\bf f}({\bf x},t)$ in the rhs of (\ref{2.1.a}) contains the
{\it density\/} of all {\it non-internal\/} forces acting on the sphere,
which we expediently split into a component due the resonators' {\it back
action\/} and an external action {\it proper\/}, which can be a GW signal,
a calibration signal, etc. Thus

\begin{equation}
  {\bf f}({\bf x},t) = {\bf f}_{\rm resonators}({\bf x},t) +
  {\bf f}_{\rm external}({\bf x},t)    \label{2.2}
\end{equation}

Finally, $\xi_a(t)$ in the rhs of (\ref{2.1.b}) is the force per unit mass
(acceleration) acting on the $a\/$-th resonator due to {\it external\/}
agents.

Since we are making the hypothesis that the resonators are point masses the
following holds:

\begin{equation}
    {\bf f}_{\rm resonators}({\bf x},t) =
    \sum_{a=1}^J M_a\Omega_a^2\,\delta^{(3)}({\bf x}-{\bf x}_a)\,
    \left[\,z_a(t)-u_a(t)\right]\,{\bf n}_a
    \label{2.3}
\end{equation}

where $\delta^{(3)}\/$ is the three dimensional Dirac density function.

The {\it external\/} forces we shall be considering in this paper will be
{\it gravitational wave\/} signals (of course!) and a simple calibration
signal, a perpendicular {\it hammer stroke\/}. GW driving terms, we recall
from (\ref{1.1}), can be written

\begin{equation}
   {\bf f}_{\rm GW}({\bf x},t) = {\bf f}^{(00)}({\bf x})\,g^{(00)}(t)\ +\ 
    \sum_{m=-2}^2\,{\bf f}^{(2m)}({\bf x})\,g^{(2m)}(t)    \label{2.4}
\end{equation}
for a general {\it metric\/} wave ---see \cite{lobo} for explicit
formulas and technical details. While the spatial coefficients
${\bf f}^{(lm)}({\bf x})$ are pure {\it form factors\/} associated to
the {\it tidal\/} character of a GW excitation, it is the time dependent
factors $g^{(lm)}(t)$ which carry the specific information on the
incoming GW. The purpose of a GW detector is to determine the latter
coefficients on the basis of suitable measurements.

If a GW sweeps the observatory then the resonators themselves will also be
affected, of course. They will be driven, relative to the sphere's centre,
by a tidal acceleration which, since they only move radially, is given by

\begin{equation}
   \xi_a^{\rm GW}(t) = c^2\,R_{0i0j}(t)\,x_{a,i}n_{a,j}\ ,
   \qquad a=1,\ldots,J      \label{c.1}
\end{equation}
where $R_{0i0j}(t)$ are the ``electric'' components of the GW Riemann
tensor at the centre of the sphere, and $R\/$ is the sphere's radius so
that, clearly, ${\bf x}_a$\,=\,$R{\bf n}_a$. It may now be recalled from
reference \cite{lobo} that

\begin{equation}
   c^2\,R_{0i0j}(t) =
   \sum_{\stackrel{\scriptstyle l=0\ {\rm and}\ 2}{m=-l,...,l}}\,
   E_{ij}^{(lm)}\,g^{(lm)}(t)          \label{c.2}
\end{equation}

where $E_{ij}^{(lm)}$ is a set of 6 (constant) symmetric matrices which
verify\footnote{
$Y_{lm}({\bf n})$ are spherical harmonics \protect\cite{Ed60}.}

\begin{equation}
   E_{ij}^{(lm)}\,n_i n_j = Y_{lm}({\bf n})\ ,\qquad
   l=0,2\ ,\ \ m=-l,\ldots,l     \label{c.3}
\end{equation}

Finally thus:

\begin{equation}
   \xi_a^{\rm GW}(t) = R\,
   \sum_{\stackrel{\scriptstyle l=0\ {\rm and}\ 2}{m=-l,...,l}}\,
   Y_{lm}({\bf n}_a)\,g^{(lm)}(t)\ ,\qquad a=1,\ldots,J   \label{c.4}
\end{equation}

We shall also be eventually considering in this paper the response of the
system to a particular {\it calibration\/} signal, consisting in a hammer
stroke with intensity ${\bf f}_0$, delivered perpendicularly to the
sphere's surface at point ${\bf x}_0$:

\begin{equation}
   {\bf f}_{\rm stroke}({\bf x},t) = {\bf f}_0\,
   \delta^{(3)}({\bf x}-{\bf x}_0)\,\delta(t)   \label{2.5}
\end{equation}

which we have modeled as an impulsive force in both space and time
variables. Unlike GW tides, a hammer stroke will be applied on the sphere's
surface, so it may have no {\it direct\/} effect on the resonators. In
other words,

\begin{equation}
   \xi_a^{\rm stroke}(t) = 0\ ,\qquad a=1,\ldots,J    \label{c.5}
\end{equation}

Our fundamental equations thus read

\begin{mathletters}
\label{2.6}
\begin{eqnarray}
    \varrho \frac{\partial^2 {\bf u}}{\partial t^2} & = & \mu\nabla^2 {\bf u}
    + (\lambda+\mu)\,\nabla(\nabla{\bf\cdot}{\bf u}) +
    \nonumber \\
    & & \sum_{b=1}^J M_b\Omega_b^2\,\delta^{(3)}({\bf x}-{\bf x}_b)\,
    \left[z_b(t)-u_b(t)\right]\,{\bf n}_b
    + {\bf f}_{\rm external}({\bf x},t)    \label{2.6.a}  \\*[0.7 em]
    \ddot{z}_a(t) & = & -\Omega_a^2\,
    \left[z_a(t)-u_a(t)\right] + \xi_a(t)\ ,
    \qquad a=1,\ldots,J    \label{2.6.b}
\end{eqnarray}
\end{mathletters}

where ${\bf f}_{\rm external}({\bf x},t)$ will be given by either (\ref{2.4})
or (\ref{2.5}), as the case may be. Likewise, $\xi_a(t)$ will be given by
(\ref{c.4}) or (\ref{c.5}), respectively. The remainder of this paper will
be concerned with finding solutions to the system of coupled differential
equations (\ref{2.6}), and with their meaning and consequences.

\section{Green function formalism}

An elegant and powerful method to solve equations (\ref{2.6}) is the Green
function formalism. It so happens that, in all instances of our concern here,
the force density ${\bf f}({\bf x},t)$ of equation (\ref{2.2}) is of the
{\it separable type\/}, i.e., it can be written as a sum of products of
a function of the space variables {\bf x} times a function of the time
variable $t\/$. Direct inspection of equations (\ref{2.3})-(\ref{2.5})
readily shows that this is {\it always\/} the case. We thus have, generically,

\begin{equation}
  {\bf f}({\bf x},t) = \sum_\alpha\,{\bf f}^{(\alpha)}({\bf x})\,
   g^{(\alpha)}(t)   \label{3.1}
\end{equation}
where $\alpha\/$ is a suitable label. We recall from reference \cite{lobo}
that, in such circumstances, a formal solution can be written down for
equation (\ref{2.1.a}) in terms of a {\it Green function integral\/},
whereby the following orthogonal series expansion obtains:

\begin{equation}
   {\bf u}({\bf x},t) = \sum_\alpha\sum_N\,\omega_N^{-1}\,f_N^{(\alpha)}\,
   {\bf u}_N({\bf x})\,g_N^{(\alpha)}(t)      \label{3.2}
\end{equation}

where

\begin{mathletters}
\label{3.3}
\begin{eqnarray}
  f_N^{(\alpha)} & \equiv & \frac{1}{\cal M}\,\int_{\rm Sphere}
  {\bf u}_{N}^*({\bf x})\cdot{\bf f}^{(\alpha)}({\bf x})\,d^3x
  \label{3.3.a} \\[0.5 em]
  g_N^{(\alpha)}(t) & \equiv & \int_0^t g^{(\alpha)}(t')\,\sin\omega_N (t-t')
  \,dt'   \label{3.3.b}
\end{eqnarray}
\end{mathletters}

Here, $\omega_N\/$ and ${\bf u}_N({\bf x})$ are the eigenfrequencies and
associated normalised wavefunctions of the free sphere ---see again
\cite{lobo} for a comprehensive characterisation. Also, $N\/$ is an
abbreviation for a multiple index $\{nlm\}$. We quote the result of a few
explicit calculations which will be useful later on:

\begin{mathletters}
\label{3.4}
\begin{eqnarray}
 f_{{\rm resonators,} N}^{(a)} & = & \frac{M_a}{\cal M}\,\Omega_a^2\,\,
  {\bf n}_a\!\cdot\!{\bf u}_N^*({\bf x}_a)\ \ ,\qquad a=1,\ldots,J
  \label{3.4.a} \\[0.5 em]
 f_{{\rm GW,}N}^{(l'm')} & = & a_{nl}\,\delta_{ll'}\,\delta_{mm'}\ \ ,
  \qquad N\equiv\{nlm\}\ ,\ \ l'=0,2\ ,\ \ m'=-l',\ldots,l'
  \label{3.4.b} \\[0.5 em]
 f_{{\rm stroke,}N} & = &{\cal M}^{-1}\,
                        {\bf f}_0\!\cdot\!{\bf u}_N^*({\bf x}_0)
  \label{3.4.c}
\end{eqnarray}
\end{mathletters}

where the coefficients $a_{nl}\/$ in (\ref{3.4.b}) are overlapping integrals
of ${\bf f}^{(lm)}({\bf x})$ across the volume of the sphere \cite{foot1},
and

\begin{mathletters}
\label{3.5}
\begin{eqnarray}
 g_{{\rm resonators,} N}^{(a)}(t) & = & \int_0^t\left[z_a(t')-u_a(t)
   \right]\,\sin\omega_N(t-t')\,dt'
   \ \ ,\qquad a=1,\ldots,J  \label{3.5.a}  \\[0.5 em]
 g_{{\rm GW,}N}^{(lm)}(t) & = & \int_0^t g^{(lm)}(t')\,
   \sin\omega_N(t-t')\,dt'  \label{3.5.b}  \\[0.7 em]
 g_{{\rm stroke,}N}(t) & = & \sin\omega_Nt  \label{3.5.c}
\end{eqnarray}
\end{mathletters}

These expressions can now be substituted into equation (\ref{3.2}) to obtain
the following equivalent set of equations of motion:

\begin{mathletters}
\label{3.6}
\begin{eqnarray}
 {\bf u}({\bf x},t) & = & \sum_N\,\omega_N^{-1}\,{\bf u}_N({\bf x})\,
  \left\{\sum_{b=1}^J\,\frac{M_b}{\cal M}\,\Omega_b^2\,
  \left[{\bf n}_b\!\cdot\!{\bf u}_N^*({\bf x}_b)\right]\,
  g_{{\rm resonators,} N}^{(b)}(t) + \sum_\alpha
  f_{{\rm external,} N}^{(\alpha)}\,g_{{\rm external,} N}^{(\alpha)}(t)
  \right\}	\label{3.6.a} \\
  \ddot{z}_a(t) & = & -\Omega_a^2\,
  \left[z_a(t)-u_a(t)\right] + \xi_a(t)\ ,
  \qquad a=1,\ldots,J    \label{3.6.b}
\end{eqnarray}
\end{mathletters}

We now specify {\bf x}\,=\,${\bf x}_a\/$ in (\ref{3.6.a}) and multiply both
sides of the equation by ${\bf n}_a\/$, thus finding

\begin{mathletters}
\label{3.7}
\begin{eqnarray}
  u_a(t) & = & u_a^{\rm external}(t) + \sum_{b=1}^J\,\eta_b\,\int_0^t
  K_{ab}(t-t')\,\left[\,z_b(t')-u_b(t')\right]\,dt'  \label{3.7.a}\\
  \ddot{z}_a(t) & = & -\Omega_a^2\,\left[\,z_a(t)-u_a(t)\right]+\xi_a(t)
  \ , \qquad a=1,\ldots,J    \label{3.7.b}
\end{eqnarray}
\end{mathletters}

where $u_a^{\rm external}(t)$\,$\equiv$\,
${\bf n}_a\!\cdot\!{\bf u}^{\rm external}({\bf x}_a,t)$, and

\begin{equation}
   {\bf u}^{\rm external}({\bf x},t) = \sum_\alpha\sum_N\,\omega_N^{-1}\,
    f_{{\rm external,}N}^{(\alpha)}\,{\bf u}_N({\bf x})\,
    g_{{\rm external,}N}^{(\alpha)}(t)      \label{3.9}
\end{equation}

is the sphere's response to an {\it external force in the absence of
resonators\/}. The form of (\ref{3.9}) is given in reference \cite{lobo}
for a generic metric GW signal and for a hammer stroke. The {\it kernel
matrix\/} $K_{ab}(t)$ in (\ref{3.7.b}) is the following weighted sum of
diadic products of wavefunctions:

\begin{equation}
  K_{ab}(t) = \Omega_b^2\,\sum_N\,\omega_N^{-1}\,
  \left[{\bf n}_b\!\cdot\!{\bf u}_N^*({\bf x}_b)\right]
  \left[{\bf n}_a\!\cdot\!{\bf u}_N({\bf x}_a)\right]\,\sin\omega_Nt
  \label{3.10}
\end{equation}

Finally, we have defined the mass ratios of the resonators to the entire
sphere

\begin{equation}
  \eta_b\equiv \frac{M_b}{\cal M}\ ,\qquad b=1,\ldots,J   \label{3.11}
\end{equation}

which will be {\it small parameters\/} in a real device.

Equations (\ref{3.7}) are a set of integro-differential equations for the
radial deformations of the sphere, $u_a(t)$, and those of the resonators,
$z_a(t)$. If we solve them then we obtain at once the complete solution
to our general problem by direct substitution of these quantities into
(\ref{3.5.a}), then in (\ref{3.6.a}). But before going into the technical
details of the solution process let us briefly pause for a qualitative
inspection. 

Equation (\ref{3.7.a}) shows that the sphere's deformations $u_a(t)$ are
made up of two contributions: one due to the action of {\it external\/}
agents (GWs or other), contained in $u_a^{\rm external}(t)$, and another
one due to coupling to the resonators. The latter is commanded by the
small parameters $\eta_b\/$, and correlates to {\it all\/} of the sphere's
spheroidal eigenmodes through the kernel matrix $K_{ab}(t)$. This has
consequences for GW detectors, for even though GWs only couple to
quadrupole and monopole\footnote{
Monopole modes only exist in scalar-tensor theories of gravity, such as e.g.
Brans--Dicke \protect\cite{bd61}; General Relativity does not belong in this
category.}
spheroidal modes of the {\it free\/} sphere \cite{wp77,lobo,bian},
attachement of resonators causes, as we see, a well defined amount of
energy to be transferred from these into other modes of the antenna, and
conversely, these modes back-act on the former. As we shall shortly prove,
such effects can be minimised by suitable {\it tuning\/} of the resonators'
frequencies, but outright neglection of them results in inaccurate
conclusions about the system dynamics.

\subsection{Laplace transform domain equations}

We now take up the problem of solving equations (\ref{3.7}). Equation
(\ref{3.7.a}) is an integral equation belonging in the general category
of Volterra equations \cite{tricomi}, but a series solution to it in
ascending powers of $\eta_b\/$ by iterative substitution of $u_b(t)$ into
the kernel integral is not viable due to the {\it dynamical\/} contribution
of $z_b(t)$, which is in turn governed by the {\it differential\/}
equation (\ref{3.7.b}). A better suited method to address this
{\it integro-differential\/} system is to Laplace-transform equations
(\ref{3.7}). We denote the Laplace transform of a generic function of
time $f(t)$ with a {\it caret\/} (\,$\hat{}$\,) on its symbol, e.g.,

\begin{equation}
  \hat{f}(s) \equiv \int_0^\infty f(t)\,e^{-st}\,dt	\label{3.12}
\end{equation}

and make the assumption that the system is at rest before an instant of
time, $t\/$\,=\,0, say, or

\begin{equation}
  {\bf u}({\bf x},0)={\bf\dot u}({\bf x},0)=z_a(0)=\dot z_a(0) = 0
  \label{3.14}
\end{equation}

Equations (\ref{3.7}) are then recast in the equivalent form

\begin{mathletters}
\label{3.13}
\begin{eqnarray}
    \hat{u}_a(s) & = & \hat{u}_{a}^{\rm external}(s)
    - \,\sum_{b=1}^J \eta_b\,\hat K_{ab}(s)\,
    \left[\hat z_b(s)-\hat{u}_b(s)\right]
    \label{3.13.a}   \\*[0.7 em]
    s^2\,\hat{z}_a(s) & = & -\Omega_a^2\,
    \left[\hat z_a(s)-\hat{u}_a(s)\right] + \hat\xi_a(s)\ ,
    \qquad a=1,\ldots,J
    \label{3.13.b}
\end{eqnarray}
\end{mathletters}
 
for which use has been made of the {\it convolution theorem\/} for Laplace
transforms\footnote{
This theorem states, it is recalled, that the Laplace transform of the
convolution product of two functions is the arithmetic product of their
respective Laplace transforms.}.
A further simplification is accomplished if we consider that we shall
in practice be only concerned with the {\it measurable\/} quantities

\begin{equation}
  q_a(t)\equiv z_a(t)-u_a(t) \ ,\qquad  a=1,\ldots,J	 \label{3.15}
\end{equation}

representing the resonators' actual elastic deformations ---cf.\ Figure
\ref{fig1}. It is readily seen that these verify the following:

\begin{equation}
  \sum_{b=1}^J \left[\delta_{ab} + \eta_b\,\frac{s^2}{s^2+\Omega_a^2}\,
  \hat K_{ab}(s)\right]\,\hat q_b(s) = -\frac{s^2}{s^2+\Omega_a^2}\,
  \hat u_a^{\rm external}(s) + \frac{\hat\xi_a(s)}{s^2+\Omega_a^2}\ ,
  \qquad a=1,\ldots,J	\label{3.16}
\end{equation}

where

\begin{equation}
  \hat u_a^{\rm external}(s) = \sum_\alpha\,\left\{\sum_N\,
   \frac{1}{s^2+\omega_N^2}\,f_{{\rm external,}N}^{(\alpha)}\,
   \left[{\bf n}_a\!\cdot\!{\bf u}_N({\bf x}_a)\right]\right\}\,
   \hat g_{\rm external}^{(\alpha)}(s)	  \label{3.165}
\end{equation}

and

\begin{equation}
  \hat K_{ab}(s) = \sum_N\,\frac{\Omega_b^2}{s^2+\omega_N^2}\,
   \left[{\bf n}_b\!\cdot\!{\bf u}_N^*({\bf x}_b)\right]
   \left[{\bf n}_a\!\cdot\!{\bf u}_N({\bf x}_a)\right]
   \label{3.17}
\end{equation}

which ensue directly from (\ref{3.9}), (\ref{3.10}) and the definition
(\ref{3.12}).

Equations (\ref{3.16}) constitute a significant simplification of the
original problem, as they are a set of just $J\/$ {\it algebraic\/} rather
than integral or differential equations. We must solve them for the unknowns
$\hat q_a(s)$, then perform {\it inverse Laplace transforms\/} to revert to
$q_a(t)$. But we note that Equations (\ref{3.16}) and (\ref{3.165})
indicate that the Laplace transform functions we shall be concerned with
are of the {\it rational\/} class, i.e., they are {\it quotients of
polynomials\/} in $s\/$. This greatly facilitates the latter step
(Laplace transform inversion), as it can in this case be obtained by
the {\it calculus of residues\/}~\cite{porter}.

Determination of the {\it poles\/} of $\hat q_a(s)$ is thus required in
the first place. Clearly, poles correspond to those values of $s\/$ for
which the matrix in square brackets in the lhs of (\ref{3.16}) is
{\it singular\/}\footnote{
I.e., non-invertible.}
or, equivalently, to the zeroes of its determinant:

\begin{equation}
  \Delta(s)\equiv\det\left[\delta_{ab} + \eta_b\,
  \frac{s^2}{s^2+\Omega_a^2}\,\hat K_{ab}(s)\right]= 0\ ,
  \qquad {\rm Poles}	\label{3.18}
\end{equation}

As is well known, the {\it imaginary parts\/} of the {\it poles\/} are
the system {\it characteristic frequencies\/}, or resonances
\cite{hels}; {\it residues\/} at such poles determine the specific
weight of the respectively associated modes in the system response to
a given external agent.

The procedure's general guidelines to solve the problem are thus clear-cut.
The details of its actual implementation are however not obvious, so we now
come to them. We begin with the simplest case, i.e., a perfect sphere with
perfectly tuned identical resonators.

\section{Perfect sphere, identical resonators, ideal tuning}

We first consider this highly idealised situation as a suitable starting
point to address more realistic practical instances, which we naturally
do {\it not\/} expect to depart excessively from that ideality, anyway.
Moreover this simpler analysis will be very interesting {\it conceptually\/},
as all the major (coarse grain) characteristics of the system will emerge
out of it.

In a perfect sphere the frequencies of the eigenmodes can be classified as
$l\/$-pole series of ascending harmonics, each frequency in a given series
being (2$l\/$+1)-fold degenerate. GWs exclusively couple to
{\it spheroidal\/} eigenmodes \cite{bian}, and our resonators are assumed
to be sensitive to {\it radial\/} deformations of the sphere, so only
{\it spheroidal frequencies\/} will concern us here. Following the notation
of reference \cite{lobo}, we denote by $\omega_{nl}\/$ the $n\/$-th harmonic
of the $l\/$-th $l\/$-pole series, whose associated 2$l\/$+1 eigenfunctions
have the following radial projections:

\begin{equation}
 {\bf n}\!\cdot\!{\bf u}_{nlm}({\bf x}) = A_{nl}(r)\,Y_{lm}(\theta,\varphi)
 \label{4.1}
\end{equation}

where $Y_{lm}(\theta,\varphi)$ are spherical harmonics \cite{Ed60}, and
$A_{nl}(r)$ are given in \cite{lobo}. Degeneracy of the eigenfrequencies
$\omega_{nl}$ enables direct summation over the degeneracy index $m\/$ in
(\ref{3.17}), so that

\begin{equation}
  \hat K_{ab}(s) = \sum_{nl}\,\frac{\Omega_b^2}{s^2+\omega_{nl}^2}\,
   \left|A_{nl}(R)\right|^2\,\frac{2l+1}{4\pi}\,
   P_l({\bf n}_a\!\cdot\!{\bf n}_b) \equiv
   \sum_{nl}\,\frac{\Omega_b^2}{s^2+\omega_{nl}^2}\,\chi_{ab}^{(nl)}
   \label{4.2}
\end{equation}

for a {\it perfectly symmetric sphere\/}. Here, use has been made of the
summation formula (\ref{A.1}) ---see Appendix below.

Our next assumption is that all the resonators are {\it identical\/}, or

\begin{equation}
  \eta_1=\,\ldots\,=\eta_J\equiv\eta\ ,\qquad
  \Omega_1=\,\ldots\,=\Omega_J\equiv\Omega
  \label{4.5}
\end{equation}

The fundamental idea behind using resonators is to have them tuned to one
of the frequencies of the sphere's spectrum, so we now make our third
hypothesis:

\begin{equation}
  \Omega = \omega_{n_0l_0}	\label{4.6}
\end{equation}

In a GW detector it will only make sense to choose $l_0$\,=\,0 or
$l_0$\,=\,2, of course, and have $n_0$ refer to the first or perhaps
second harmonic \cite{clo}. We keep the generic expression (\ref{4.6}) for
the time being to encompass all the possibilities within a single formalism,
and to make room for calibration signals, too.

Based on the above three hypotheses ---i.e., perfect spherical symmetry
plus equations (\ref{4.5}) and (\ref{4.6})---, we can now rewrite equation
(\ref{3.17}):

\begin{equation}
 \sum_{b=1}^J\,\left[\delta_{ab} + \eta\,\sum_{nl}\,
   \frac{\Omega^2s^2}{(s^2+\Omega^2)(s^2+\omega_{nl})^2}\,\chi_{ab}^{(nl)}
   \right]\,\hat q_b(s) = -\frac{s^2}{s^2+\Omega^2}\,
   \hat u_a^{\rm external}(s) + \frac{\hat\xi_a(s)}{s^2+\Omega_a^2}\ ,
   \qquad  (\Omega = \omega_{n_0l_0})
   \label{4.8}
\end{equation}

Equation (\ref{3.18}) for the system resonances can also be recast in a
more convenient form:

\begin{equation}
  \Delta(s) \equiv \det\,\left[\delta_{ab} + \eta\,\frac{\Omega^2s^2}
   {(s^2+\Omega^2)^2}\,\chi_{ab}^{(n_0l_0)} + \eta\,\sum_{nl\neq n_0l_0}\,
   \frac{\Omega^2s^2}{(s^2+\Omega^2)(s^2+\omega_{nl}^2)}\,\chi_{ab}^{(nl)}
   \right] = 0          \label{4.9}
\end{equation}

To find an {\it analytic\/} expression for the roots of (\ref{4.9}) is an
impossible task, so we resort to a {\it perturbative expansion\/} in the
small ``coupling constant'' $\eta\/$. As we shall now see, much light is
shed onto the physics of the problem as we proceed with that expansion.

\section{Frequency spectrum}

Let us formally state the {\it essential\/} fact that the resonators are
much less masssive than the sphere, which is the basis for {\it all\/}
the considerations from now on in this paper:

\begin{equation}
  \eta\ll 1	\label{4.10}
\end{equation}

It is clear from the structure of equation (\ref{4.9}) that vanishing of
$\Delta(s)$ will require $s\/$ to be such that the denominators in the
fractions within square brackets there be proportional to $\eta\/$;
otherwise it will not be possible to have $\Delta(s)$\,=\,0 for
{\it arbitrarily small\/} values of $\eta\/$, since $\delta_{ab}$ is of
course a regular matrix. But this points to a sharp distinction between roots
which are close to $s\/$\,=\,$\pm i\Omega$\,=\,$\pm i\omega_{n_0l_0}$, and
roots which are close to $s\/$\,=\,$\pm i\omega_{nl}$ (${nl\neq n_0l_0}$).
The former's {\it degree of proximity\/} to $\pm i\Omega$ is of order
$\eta^{1/2}$, while the latter's proximity to $\pm i\omega_{nl}$ is of
order $\eta\/$, instead. We thus correspondingly distinguish the following
two categories of roots:

\begin{mathletters}
\label{4.11}
\begin{eqnarray}
    s_0^2 & = & -\Omega^2\,\left(1 + \chi_\frac{1}{2}\,\eta^{1/2}
    + \chi_1\,\eta + \ldots\right)  \qquad (\Omega=\omega_{n_0l_0})
    \label{4.11.a}   \\
    s_{nl}^2 & =& -\omega_{nl}^2\,\left(1 + b_1^{(nl)}\,\eta +
	b_2^{(nl)}\,\eta^2 + \ldots\right) \qquad ({nl\neq n_0l_0})
    \label{4.11.b}
\end{eqnarray}
\end{mathletters}

The coefficients $\chi_\frac{1}{2}$, $\chi_1$,... and $b_1^{(nl)}$,
$b_1^{(nl)}$,... can be calculated recursively, starting form the first.
As we now show, these two categories of roots present qualitatively
different characteristics. We consider them separately.

\subsection{Roots near $\Omega$}

Upon substitution of (\ref{4.11.a}) into (\ref{4.9}) it is readily seen
that $\chi_\frac{1}{2}$ is a solution to the algebraic equation

\begin{equation}
  \det\left[\delta_{ab} - \frac{1}{\chi_\frac{1}{2}^2}\,
  \chi_{ab}^{(n_0l_0)}\right] = 0
  \label{5.1}
\end{equation}

\begin{figure}[htb]
\psfig{file=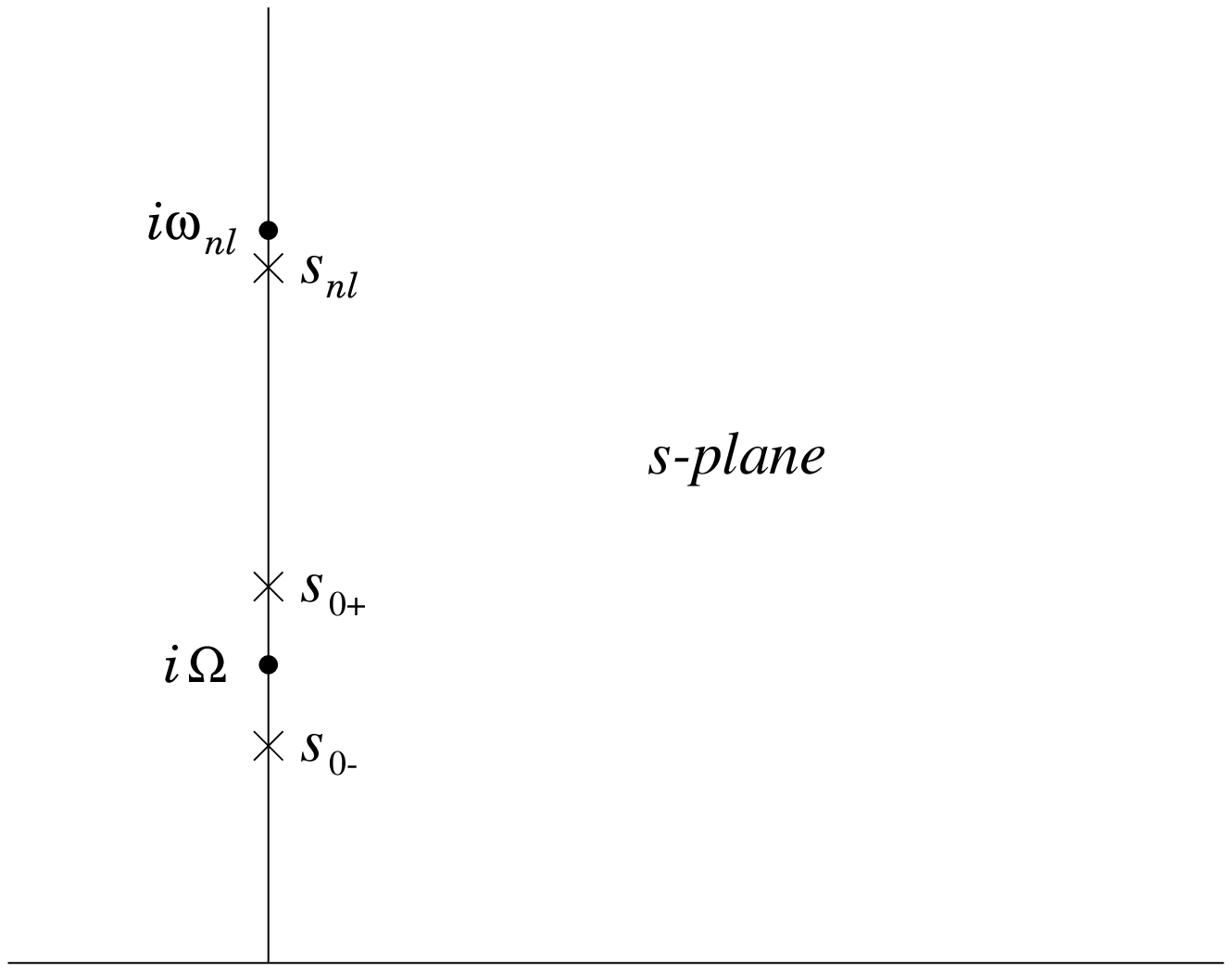,height=16.4cm,width=12cm,rheight=6.8cm,bbllx=-2cm,bblly=-5.5cm,bburx=15.2cm,bbury=21.5cm}
\caption{Qualitative situation of the roots in the complex $s\/$-plane.
Dots correspond to the {\it free\/} sphere's frequencies, while crosses
correspond to {\it coupled\/} system frequencies. Roots near the tuning
frequency $\Omega$ come in pairs (we draw just one, corresponding to one
single resonator, to avoid unnecessary complication at this stage), while
roots near other (higher) free sphere frequencies are {\it downshifted\/}
rather than split, though the effect is altogether weaker in this case
---see equation (\protect\ref{4.11.b}).   \label{fig2}}
\end{figure}

So $\chi_\frac{1}{2}^2$ are the {\it eigenvalues\/} of the matrix
$\chi_{ab}^{(n_0l_0)}$, which happens to be non-negative definite ---see
Appendix. There are of course $J\/$ of these, some of which may be
repeated ({\it degenerate\/}), or null. This means that we get $J\/$
{\it pairs of roots\/} of the type $s_0$ in (\ref{4.11.a}) lying on the
imaginary axis of the complex $s\/$-plane {\it symmetrically\/} (to order
$\eta^{1/2}$) around $\pm i\Omega$  ---see Figure \ref{fig2}. The system
resonant frequencies can thus be represented by the pairs

\begin{equation}
  \omega_{a\pm}^2 = \Omega^2\,\left(1\pm\sqrt{\frac{2l+1}{4\pi}}\,
  \left|A_{n_0l_0}(R)\right|\,\zeta_a\,\eta^{1/2}\right) + O(\eta)\ ,
  \qquad a=1,\ldots,J
  \label{5.2}
\end{equation}
where $\zeta_a^2\/$ are the {\it eigenvalues\/} of the matrix
$P_{l_0}({\bf n}_a\!\cdot\!{\bf n}_b)$, and $O(\eta)$ stands for the
higher order terms in (\ref{4.11.a}).

Equation (\ref{5.2}) very much reminds us of a familiar result in
cylindrical bar theory \cite{pia}: attachement of a resonator to the bar's
end face causes the latter's characteristic resonance frequency to split
up into a symmetric pair around the original value, the amount of relative
shift being proportional to $\eta^{1/2}$. As we now see, in a spherical
body the number of frequency pairs equals the number of resonators, it is
of the same order of magnitude ($\eta^{1/2}$), and is precisely controlled
by the geometry dependent eigenvalues $\zeta_a\/$.

It must however be cautioned that (\ref{5.2}) {\it is not an exact\/}
result, but the {\it dominant approximation\/} in an ascending series in
powers of $\eta^{1/2}$. If better accuracy is needed then higher terms in
(\ref{4.11.a}) must be calculated. For example, the {\it next order\/}
correction to $s_0^2$ in (\ref{4.11.a}) is

\begin{equation}
  \chi_{1,a} = \frac{1}{2}\,\left[\chi_{\frac{1}{2},a}^2 -
  \sum_{nl\neq n_0l_0}\,\frac{\Omega^2}{\omega_{nl}^2-\Omega^2}\,
  \tilde\chi_{aa}^{(nl)}\right]\ , \qquad a=1,\ldots,J
  \label{5.3}
\end{equation}
corresponding to the $a\/$-th root of the previous order, equation
(\ref{5.1}). In this formula we have $\tilde\chi_{aa}^{(nl)}$, which stands
for the $a\/$-th diagonal element of the matrix $\chi_{ab}^{(nl)}$ in the
(abstract) system of axes in which $\chi_{ab}^{(n_0l_0)}$ is diagonal.

The second order corrections (\ref{5.3}) are seen to be equal for the two
members of the frequency pairs (\ref{5.2}), as they only depend on
$\chi_\frac{1}{2}^2$. They thus result in {\it rigid shifts\/} for both of
them. More substantial, the actual value of these corrections involves the
{\it whole spectrum\/} of spheroidal eigenfrequencies and wavefunctions.
This should not be considered surprising on general grounds\footnote{
Second order perturbative corrections to eigenvalues often show this
feature ---recall e.g. energy level corrections in Quantum Mechanics
\protect\cite{bj}.}, but to find this result has required us to model the
system dynamics in a more elaborate conceptual structure than has been
considered so far in the literature \cite{jm95,grg,ts}. The physical
interpretation of (\ref{5.3}) is that the attachement of resonators causes
a certain level of {\it intercommunication\/} between different vibration
eigenmodes of the sphere, even if the resonators are accurately tuned to
one of the spectral frequencies. This ``cross talk'' between modes has
consequences for non-tuned modes, as we shall see in the next subsection,
as well as for tuned modes, as we see in (\ref{5.3}). The practical
relevance of these higher order corrections will naturally be dictated by
the precision of measurments in a real GW antenna. As of this date, the
experimental data available on such devices as described here is to our
knowledge limited to the {\sl TIGA\/} prototype data \cite{phd}, but the
experimental conditions reported \cite{jm97} are not sufficiently fine
tuned to guarantee that corrections of order $\eta\/$ to fit the data are
really meaningful ---see section 8 below. This is surely the reason why
errors in the formulas given by other authors for system frequencies
\cite{foot2} have passed unnoticed so far.

\subsection{Roots near $\omega_{nl}$\,$\neq$\,$\Omega$}

We now come to the other roots, equation (\ref{4.11.b}). If the latter
is substituted into (\ref{4.9}) then we readily see that $b_1^{(nl)}$
satisfies the algebraic equation

\begin{equation}
  \det\left[\frac{\Omega^2-\omega_{nl}^2}{\omega_{nl}^2}\,b_1^{(nl)}\,
  \delta_{ab} - \chi_{ab}^{(nl)}\right] = 0
  \label{5.4}
\end{equation}
which is again an eigenvalue equation, having $J\/$ solutions. Since the
eigenvalues of the matrix $\chi_{ab}^{(nl)}$ are either positive or null
(see Appendix) it follows that $b_1^{(nl)}$ either has the same sign as
$(\Omega^2-\omega_{nl}^2)$ or is zero, respectively. Thus for example, if
$\Omega$ is chosen equal to the lowest frequency of the spectrum (first
quadrupole harmonic, $\omega_{12}$) then the system frequencies close to
$\omega_{nl}$ will be slightly {\it downshifted\/} relative to their
original value $\omega_{nl}$. This is schematically represented in
Figure~\ref{fig2}.

We see again in equation (\ref{5.4}) that the presence of resonators does
indeed affect the whole spectrum of the free sphere, though with an
intensity of coupling which differs between tuned and non-tuned modes by
factors of order $\eta^{1/2}$. We shall not go into any more depth in the
analysis of these fine structure details in this paper.

\section{System response to a Gravitational Wave}

Our next concern is the actual system response when it is acted upon by an
incoming GW, i.e., which are the {\it amplitudes\/} of the excited modes,
whose frequencies we have just estimated, and how do they relate to the GW
amplitudes $g^{(lm)}(t)$. To this end we must find the inverse Laplace
transform of $\hat q_a(s)$:

\begin{equation}
   \hat q_a(s) = -\sum_{b=1}^J\,\left[\delta_{ab} + \eta\,\sum_{nl}\,
   \frac{\Omega^2s^2}{(s^2+\Omega^2)(s^2+\omega_{nl})^2}\,\chi_{ab}^{(nl)}
   \right]^{-1}\,\left(\frac{s^2}{s^2+\Omega^2}\,
   \hat u_b^{\rm GW}(s) - \frac{\hat\xi_a(s)}{s^2+\Omega_a^2}\right)
   \ ,\ \ \ a=1,\ldots,J
   \label{6.1}
\end{equation}

as follows from (\ref{4.8}), where $\hat u_b^{\rm external}(s)$ has been
substituted by $\hat u_b^{\rm GW}(s)$, which is in its turn given by

\begin{equation}
  \hat u_b^{\rm GW}(s) = 
   \sum_{\stackrel{\scriptstyle l=0\ {\rm and}\ 2}{m=-l,...,l}}\,\left(
   \sum_{n=1}^\infty\,\frac{a_{nl}\,A_{nl}(R)}{s^2+\omega_{nl}^2}\right)
   \,Y_{lm}({\bf n}_b)\,\hat g^{(lm)}(s)\ ,\qquad b=1,\ldots,J
   \label{6.2}
\end{equation}

according to (\ref{3.4.b}), (\ref{3.5.b}) and (\ref{4.1}). Clearly, from
(\ref{c.4}) we also have

\begin{equation}
   \hat\xi_a^{\rm GW}(s) = R\,
   \sum_{\stackrel{\scriptstyle l=0\ {\rm and}\ 2}{m=-l,...,l}}\,
   Y_{lm}({\bf n_a})\,\hat g^{(lm)}(s)\ ,\qquad a=1,\ldots,J   \label{c.6}
\end{equation}

In order to ease the notation we can combine (\ref{6.1}) and (\ref{6.2})
into a more compact form:

\begin{equation}
   \hat q_a(s) = \sum_{\mbox{\scriptsize $\begin{array}{c}
    l=0\ \mbox{and}\ 2 \\ m=-l,...,l \end{array}$}}\hat\Phi_a^{(lm)}(s)\,
    \hat g^{(lm)}(s)\ ,\qquad a=1,\ldots,J
    \label{6.3}
\end{equation}

with, obviously,

\begin{equation}
   \hat\Phi_a^{(lm)}(s)\equiv -\sum_{b=1}^J\,\left[\delta_{ab} +
   \eta\,\sum_{nl}\,\frac{\Omega^2s^2}{(s^2+\Omega^2)(s^2+\omega_{nl})^2}\,
   \chi_{ab}^{(nl)}\right]^{-1}\,\frac{s^2}{s^2+\Omega^2}\,\left(
   -\frac{R}{s^2} +
   \sum_{n=1}^\infty\,\frac{a_{nl}\,A_{nl}(R)}{s^2+\omega_{nl}^2}
   \right)\,Y_{lm}({\bf n}_b)      \label{6.4}
\end{equation}

We thus have, by the convolution theorem again,

\begin{equation}
   q_a(t) = \sum_{lm}\,\int_0^t\,\Phi_a^{(lm)}(t-t')\,g^{(lm)}(t')\,dt'
   \ ,\qquad a=1,\ldots,J
   \label{6.5}
\end{equation}

where $\Phi_a^{(lm)}(t)$ is the {\it inverse Laplace transform\/} of
(\ref{6.4}). Despite its complexity, $\hat\Phi_a^{(lm)}(s)$ is a
{\it rational\/} function of $s\/$, and therefore its inverse Laplace
transform can be calculated, as already advanced in section 3, by the
{\it residue theorem\/} through the formula \cite{hels}

\begin{equation}
   \Phi_a^{(lm)}(t) = 2\pi i\,\sum\,\left\{\text{residues of}\ \ 
   \left[\hat\Phi_a^{(lm)}(s)\,e^{st}\right]\ \ \text{at its poles
   in complex {\it s\/}-plane}\right\}
   \label{6.6}
\end{equation}

The {\it poles\/} in this formula happen to be {\it exclusively\/} those
of equations (\ref{4.11}), and there are {\it no\/} poles either at
$s^2$\,=\,$-\Omega^2$ or at $s^2$\,=\,$-\omega_{nl}^2$, for divergences
compensate each other in numerators and denominators in (\ref{6.4}) at
those points. Our next step is thus to calculate the corresponding
{\it residues\/}. This is a relatively simple, but considerably laborious
task, the details of which will be omitted here. We just quote the results
and directly move on to their discussion.

One can see that $\Phi_a^{(lm)}(t)$ has the following general structure:

\begin{equation}
   \Phi_a^{(lm)}(t)\propto\eta^{-1/2}\,\sum_{\zeta_c\neq 0}\,
   \left(\sin\omega_{c+}t - \sin\omega_{c-}t\right)\,\delta_{ll_0}
   + O(0)	\label{6.7}
\end{equation}
where $\omega_{c\pm}$ are given in (\ref{5.2}), and the sum above
{\it excludes\/} terms associated to null eigenvalues $\zeta_c$. The term
$O(0)$ stands for higher order corrections, which in this case happen to
be of order zero in $\eta\/$. But let us look at the important
{\it qualitative\/} consequences of (\ref{6.7}).

\begin{itemize}
 \item[\sf i)] The resonators' motions occur with a mechanical
 amplification factor of $\eta^{-1/2}$ relative to the driving GW
 amplitudes $g^{(lm)}(t)$, provided the resonator frequency $\Omega$
 is tuned to a monopole or quadrupole sphere frequency ---see the factor
 $\delta_{ll_0}$ in (\ref{6.7}). If other frequencies are chosen for
 tuning then there is no coupling to GWs to this order, but weaker by
 factors of $\eta^{1/2}$, at least.

 \item[\sf ii)] The spectral composition of these motions is dominated
 by the symmetric frequency pairs $\omega_{c\pm}$ into which the tuned
 frequency $\omega_{n_0l_0}$\,=\,$\Omega$ splits up as a consequence of
 the resonators' presence. The maximum number of such frequency pairs is
 (2$l_0$+1), {\it even if the number of resonators, J, is larger than this
 figure\/}. This is because (2$l_0$+1) is the {\it maximum\/} number of
 non-null eigenvalues $\zeta_c\/$ ---cf.\ Appendix.

 \item[\sf iii)] Higher order terms, symbolised by $O(0)$ in (\ref{6.7}),
 encompass couplings to all the non-tuned modes, whose frequencies are
 given by (\ref{4.11.b}), as well as corrections of the same order of
 magnitude in the tuned mode. The latter include in particular modes
 corresponding to a null eigenvalue $\zeta_c$, which are guaranteed to
 be present whenever $J\/$\,$>$\,2$l_0$+1. As we shall shortly see, an
 example of this is provided by the {\sl TIGA\/} layout \cite{phd}.
\end{itemize}

Higher order corrections are considerably more difficult to address than
first order, and we shall not attempt a closer approach to them in this
paper. So let us continue with the investigation of the system response
to lowest order. Given the above discussion, it is expedient to recast
(\ref{6.3}) in the form

\begin{equation}
    \hat q_a(s) = \eta^{-1/2}\,\sum_{l,m}\,\hat\Lambda_a^{(lm)}(s;\Omega)\,
    \hat g^{(lm)}(s) + O(0) \ ,\qquad a=1,\ldots,J
    \label{6.8}
\end{equation}
where $\Omega$ will be tuned to either a monopole ($\Omega$\,=\,$\omega_{n0}$)
or a quadrupole ($\Omega$\,=\,$\omega_{n2}$) frequency of the free sphere's
spheroidal spectrum. We consider the two possibilities successively.

\subsection{Monopole gravitational radiation sensing}

General Relativity, as is well known, forbids monopole GW radiation. More
general {\it metric\/} theories, e.g. Brans-Dicke \cite{bd61}, do however
predict this kind of radiation. It appears that a spherical antenna is
potentially sensitive to monopole waves, so it can serve the purpose of
thresholding, or eventually detecting them. This clearly requires that the
resonator set be tuned to a monopole harmonic of the sphere, i.e.,

\begin{equation}
   \Omega = \omega_{n0}\ ,\qquad (l_0=0)	\label{6.9}
\end{equation}
where $n\/$ tags the chosen harmonic ---most likely the first ($n\/$\,=\,1)
in a thinkable device.

Since $P_0(z)$\,$\equiv$\,1 (for all $z\/$) the eigenvalues of
$P_0({\bf n}_a\!\cdot\!{\bf n}_b)$ are, clearly,

\begin{equation}
  \zeta_1^2=J\ ,\qquad \zeta_2^2=\,\ldots\,=\zeta_J^2=0
  \label{6.10}
\end{equation}

for {\it any resonator distribution\/}. The tuned mode frequency thus splits
into a {\it single\/} strongly coupled pair:

\begin{equation}
  \omega_\pm^2 = \omega_{n0}^2\,\left(1\pm\sqrt{\frac{J}{4\pi}}\,
  \left|A_{n0}(R)\right|\,\eta^{1/2}\right) + O(\eta)\ ,
  \qquad \Omega=\omega_{n0}
  \label{6.11}
\end{equation}

The $\Lambda$-matrix of equation (\ref{6.8}) is easily seen to be in
this case

\begin{equation}
  \hat\Lambda_a^{(lm)}(s;\omega_{n0}) = (-1)^J\,\frac{a_{n0}}{\sqrt{J}}\,
  \frac{1}{2}\,\left[\left(s^2+\omega_+^2\right)^{-1} -
  \left(s^2+\omega_-^2\right)^{-1}\right]\,\delta_{l0}\,\delta_{m0}
  \label{6.12}
\end{equation}

whence the system response is

\begin{equation}
  \hat q_a(s) = \eta^{-1/2}\,\frac{(-1)^J}{\sqrt{J}}\,a_{n0}\,
  \frac{1}{2}\,\left[\left(s^2+\omega_+^2\right)^{-1} -
  \left(s^2+\omega_-^2\right)^{-1}\right]\,\hat g^{(00)}(s) + O(0)\ ,
  \qquad a=1,\ldots,J
  \label{6.13}
\end{equation}
regardless of resonator positions. The {\it overlap coefficient\/} $a_{n0}$
is calculated by means of formulas given in \cite{lobo}, and has dimensions
of length. By way of example, $a_{10}/R\/$\,$\simeq$\,0.214, and
$a_{20}/R\/$\,$\simeq$\,$-$0.038 for the first two harmonics.

A few interesting facts are displayed by equation (\ref{6.13}). First, as
we have already stressed, it is seen that if the resonators are tuned to
a monopole {\it detector\/} frequency then only monopole {\it wave
amplitudes\/} couple strongly to the system, even if quadrupole radiation
amplitudes are significantly high at the observation frequencies
$\omega_\pm\/$. Also, the amplitudes $\hat q_a(s)$ are equal for all $a\/$,
as corresponds to the spherical symmetry of monopole sphere's oscillations,
and are proportional to $J^{-1/2}$, a factor we should indeed expect as an
indication that GW {\it energy\/} is evenly distributed amongst all the
resonators. A {\it single\/} transducer suffices to experimentally
determine the only monopole GW amplitude $\hat g^{(00)}(s)$, of course,
but (\ref{6.13}) provides the system response if more than one sensor is
mounted on the antenna for whatever reasons.

\subsection{Quadrupole gravitational radiation sensing}

We now consider the more interesting case of quadrupole motion sensing.
We thus take

\begin{equation}
   \Omega = \omega_{n2}\ ,\qquad (l_0=2)	\label{6.14}
\end{equation}
where $n\/$ labels the chosen harmonic ---most likely the first
($n\/$\,=\,1) or the second ($n\/$\,=\,2) in a practical system. The
evaluation of the $\Lambda$-matrix is now considerably more involved,
yet a remarkably elegant form is found for it:

\begin{equation}
  \hat\Lambda_a^{(lm)}(s;\omega_{n2}) = (-1)^N\,\sqrt{\frac{4\pi}{5}}\,
  a_{n2}\,\sum_{b=1}^J\,\left\{\sum_{\zeta_c\neq 0}\,\frac{1}{2}\left[
  \left(s^2+\omega_{c+}^2\right)^{-1} - \left(s^2+\omega_{c-}^2\right)^{-1}
  \right]\,\frac{v_a^{(c)}v_b^{(c)*}}{\zeta_c}\right\}\,
  Y_{2m}({\bf n}_b)\,\delta_{l2}    \label{6.15}
\end{equation}

where $v_a^{(c)}$ is the $c\/$-th normalised eigenvector of
$P_2({\bf n}_a\!\cdot\!{\bf n}_b)$, associated to the {\it non-null\/}
eigenvalue $\zeta_c^2$. Let us stress once more that equation (\ref{6.15})
explicitly shows that at most 5 pairs of modes, of frequencies
$\omega_{c\pm}$, couple strongly to quadrupole GW amplitudes, {\it no matter
how many resonators in excess of 5 are mounted on the sphere\/}. The tidal
overlap coefficients $a_{2n}\/$ can also be calculated, and give for the
first two harmonics \cite{ls}

\begin{equation}
  \frac{a_{12}}{R} = 0.328\ ,\qquad\frac{a_{22}}{R} = 0.106   \label{6.16}
\end{equation}

The system response is thus

\begin{eqnarray}
  \hat q_a(s) & = & \eta^{-1/2}\,(-1)^J\,\sqrt{\frac{4\pi}{5}}\,a_{n2}\,
  \sum_{b=1}^J\,\left\{\sum_{\zeta_c\neq 0}\,\frac{1}{2}\left[
  \left(s^2+\omega_{c+}^2\right)^{-1} - \left(s^2+\omega_{c-}^2\right)^{-1}
  \right]\,\frac{v_a^{(c)}v_b^{(c)*}}{\zeta_c}\right\}\times
  \nonumber \\[0.5 em]  & & \hspace*{4 cm}
  \times\sum_{m=-2}^2\,Y_{2m}({\bf n}_b)\,\hat g^{(2m)}(s) + O(0)\ ,
  \qquad a=1\,\ldots,J	\label{6.17}
\end{eqnarray}

Equation (\ref{6.17}) is {\it completely general\/}, i.e., it is valid
for any resonator configuration over the sphere's surface, and for any
number of resonators. It describes precisely how all 5 GW amplitudes
$\hat g^{(2m)}(s)$ interact with all 5 strongly coupled system modes;
like before, {\it only\/} quadrupole {\it wave\/} amplitudes are seen
in the detector (to leading order) when $\Omega$\,=\,$\omega_{n2}$,
even if the incoming wave carries significant monopole energy at the
frequencies $\omega_{c\pm}$.

The consequences of (\ref{6.17}) are best seen in practical examples, so
we now come to a more detailed consideration of two specific resonator
distributions.

\subsection{The {\sl TIGA\/} configuration}

\begin{figure}[htb]
\psfig{file=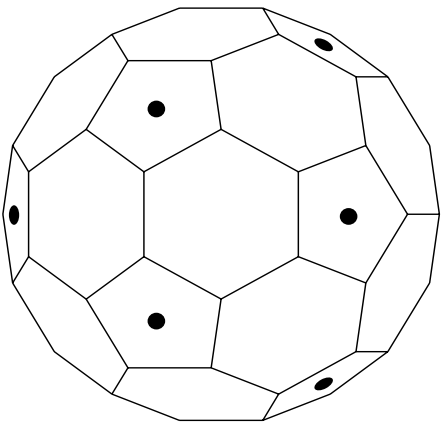,height=15cm,width=12cm,rheight=5.8cm,bbllx=-4.1cm,bblly=-3.2cm,bburx=17.6cm,bbury=25.3cm}
\caption{Schematic view of the {\sl TIGA\/}: a truncated icosahedron shape
with black dots at the resonators' posotions.
\label{ftiga}}
\end{figure}

A highly symmetric resonator layout has been proposed and experimentally
studied by Merkowitz and Johnson at {\sl LSU\/} \cite{jm93,jm95,phd},
which consists in a set of {\it six\/} transducers attached to the six
non-parallel pentagonal faces of a {\it truncated icosahedron\/}, as shown
schematically in Figure \ref{ftiga}.

It can be immediately verified that the five non-null eigenvalues of
$P_2({\bf n}_a\!\cdot\!{\bf n}_b)$ are {\it all equal\/} for this
distribution, and there is a null sixth, too:

\begin{equation}
   \zeta_{-2}^2=\,\ldots\,=\zeta_2^2=\frac{6}{5}\ ,\qquad \zeta_6^2=0
   \qquad (\text{\sl TIGA\/})	\label{6.18}
\end{equation}

This means that all five pairs of strongly coupled frequencies collapse
into a single, five-fold degenerate pair:

\begin{equation}
  \omega_\pm^2 = \omega_{n2}^2\,\left(1\pm\sqrt{\frac{3}{2\pi}}\,
  \left|A_{n2}(R)\right|\eta^{1/2}\right) + O(\eta)\ ,
  \qquad a=1,\ldots,6
  \label{6.19}
\end{equation}

Degeneracy in this layout is a consequence of its symmetry relative to the
quadrupole structure of the GW driving force and the tuned detector modes.
It is also not difficult to see that the {\sl TIGA\/} is the {\it minimal\/}
configuration with so much degeneracy, as there are no 5 resonator
configurations with equivalent symmetry. There are however other non-minimal
sets with the same degree of degeneracy ---for example 10 resonators on the
ten non-parallel faces of a regular icosahedron, etc., see e.g. \cite{grg}
for further analysis of this and other possibilities.

The normalised eigenvectors associated to the five non-vanishing eigenvalues
(\ref{6.18}), which are the only relevant ones now, are seen to be

\begin{equation}
   v_a^{(m)} = \sqrt{\frac{2\pi}{3}}\,Y_{2m}({\bf n}_a)\ ,\qquad
   m=-2,\ldots,2\ ,\ a=1,\ldots,6 \qquad (\text{\sl TIGA\/})
   \label{6.20}
\end{equation}

The system response is thus given by

\begin{equation}
  \hat q_a(s) = \eta^{-1/2}\,\sqrt{\frac{2\pi}{3}}\,a_{n2}\,
  \frac{1}{2}\left[\left(s^2+\omega_+^2\right)^{-1} -
  \left(s^2+\omega_-^2\right)^{-1}\right]\,
  \sum_{m=-2}^2\,Y_{2m}({\bf n}_a)\,\hat g^{(2m)}(s) + O(0)\ ,
  \ \ a=1\,\ldots,6	\label{6.21}
\end{equation}

This highly symmetric and remarkably simple formula was obtained for the
first time by Merkowitz and Johnson \cite{jm93,jm95}. Its scope and range
of validity, as well as its uniqueness, are now more firmly established in
the light of the present, more elaborate analysis.

Based on equations (\ref{6.21}) Merkowitz and Johnson define what they call
{\it mode channels\/}: these are linear combinations of the resonators'
readouts which directly yield the GW amplitudes $\hat g^{(2m)}(s)$ at the
frequencies $\omega_\pm$, and they are easily obtained thanks to the
orthonormality property of the eigenvectors (\ref{6.20}). They are

\begin{equation}
  \hat y^{(m)}(s)\equiv\sum_{a=1}^6\,v_a^{(m)*}\hat q_a(s) =
  \eta^{-1/2}\,a_{n2}\,
  \frac{1}{2}\left[\left(s^2+\omega_+^2\right)^{-1} -
  \left(s^2+\omega_-^2\right)^{-1}\right]\,\hat g^{(2m)}(s) + O(0)\ ,
  \ \ m=-2,\ldots,2	\label{6.22}
\end{equation}

There naturally are 5 rather than 6 mode channels ---as there are 5
quadrupole GW amplitudes---, and it is easy to implement an algorithm to
calculate them {\it on line\/} from raw detector data. So signal and
direction deconvolution methods \cite{wp77,lobo,grg} can be directly
applied to the mode channel set very advantageously. This should be
considered one more attractive property of the {\sl TIGA\/} layout since,
as we see in equation (\ref{6.17}), arbitrary resonator configurations do
not generally permit the construction of such mode channels so efficiently.

\subsection{The {\sl PHC\/} configuration}

As we have just seen the TI layout is highly symmetric, and is the minimal
set with maximum degeneracy. To accomplish this, however, 6 rather than 5
resonators are required on the sphere's surface. Since there are just 5
quadrupole GW amplitudes one may wonder whether there are alternative
layouts with {\it only\/} 5 resonators. Equation (\ref{6.17}) is completely
general, so it can be searched for an answer to this question. In reference
\cite{ls} we made a specific proposal, which we now describe in more detail.

In pursuing a search for 5 resonator sets we found that distributions having
a sphere diameter as an axes of {\it pentagonal symmetry\/}\footnote{
By this we mean resonators are placed along a {\it parallel\/} of the
sphere every 72$^\circ$.}
exhibit a rather appealing structure. More specifically, let the resonators
be located at the spherical positions

\begin{equation}
   \theta_a  = \alpha \qquad (\text{all } a)\ ,\qquad
   \varphi_a = (a-1)\,\frac{2\pi}{5}\ ,\qquad a=1,\ldots,5
\end{equation}

The eigenvalues and eigenvectors of $P_2({\bf n}_a\!\cdot\!{\bf n}_b)$ are
then

\begin{mathletters}
\label{6.24}
\begin{eqnarray}
   & \zeta_0^2 = \frac{5}{4}\,\left(3\,\cos^2\alpha-1\right)^2\ ,\qquad
       \zeta_1^2 = \zeta_{-1}^2 = \frac{15}{2}\,\sin^2\alpha\,\cos^2\alpha
       \ ,\qquad\zeta_2^2 = \zeta_{-2}^2 = \frac{15}{8}\,\sin^4\alpha  &
	\label{6.24.a} \\[1 em]
   & v_a^{(m)} = \sqrt{\frac{4\pi}{5}}\,\zeta_m^{-1}\,Y_{2m}({\bf n}_a)\ ,
       \qquad m=-2,\ldots,2\ ,\ \ a=1,\ldots,5	&
	\label{6.24.b}
\end{eqnarray}
\end{mathletters}

so the $\Lambda$-matrix is also considerably simple in structure in this
case:

\begin{equation}
  \hat\Lambda_a^{(lm)}(s;\omega_{n2}) = -\sqrt{\frac{4\pi}{5}}\,a_{n2}\,
  \zeta_m^{-1}\,\frac{1}{2}\left[\left(s^2+\omega_{m+}^2\right)^{-1} -
  \left(s^2+\omega_{m-}^2\right)^{-1}\right]\,Y_{2m}({\bf n}_a)\,\delta_{l2}
  \qquad ({\sl PHC})     \label{6.25}
\end{equation}

where we have used the obvious notation

\begin{equation}
  \omega_{m\pm}^2 = \omega_{n2}^2\,\left(1\pm\sqrt{\frac{5}{4\pi}}\,
  \left|A_{n2}(R)\right|\,\zeta_m\,\eta^{1/2}\right) + O(\eta)\ ,
  \qquad m=-2,\ldots,2
  \label{6.26}
\end{equation}

As we see from these formulas, the {\it five\/} expected pairs of
frequencies actually reduce to {\it three\/}, so pentagonal distributions
keep a certain degree of degeneracy, too. The most important distinguishing
characteristic of the general {\it pentagonal\/} layout is best displayed
by the explicit system response:

\begin{eqnarray}
  & \hat q_a(s) = -\eta^{-1/2}\,\sqrt\frac{4\pi}{5}\,a_{n2} & \left\{\,
  \frac{1}{2\zeta_0}\left[
  \left(s^2+\omega_{0+}^2\right)^{-1} - \left(s^2+\omega_{0-}^2\right)^{-1}
  \right]\,Y_{20}({\bf n}_a)\,\hat g^{(20)}(s)\right. \nonumber \\
  & & + \;\frac{1}{2\zeta_1}\left[
  \left(s^2+\omega_{1+}^2\right)^{-1} - \left(s^2+\omega_{1-}^2\right)^{-1}
  \right]\,\left[
     Y_{21}({\bf n}_a)\,\hat g^{(11)}(s) +
     Y_{2-1}({\bf n}_a)\,\hat g^{(1\,-1)}(s)\right] \label{6.27}  \\
  & & + \left.\frac{1}{2\zeta_2}\left[
  \left(s^2+\omega_{2+}^2\right)^{-1} - \left(s^2+\omega_{2-}^2\right)^{-1}
  \right]\,\left[
     Y_{22}({\bf n}_a)\,\hat g^{(22)}(s) +
     Y_{2-2}({\bf n}_a)\,\hat g^{(2\,-2)}(s)\right]\right\}
  \nonumber
\end{eqnarray}

This equation indicates that {\it different wave amplitudes selectively
couple to different detector frequencies\/}. This should be considered a
very remarkable fact, for it thence follows that simple inspection of the
system readout {\it spectrum\/}\footnote{
In a noiseless system, of course}
immediately reveals whether a given wave amplitude $\hat g^{2m}(s)$ is
present in the incoming signal or not.

Pentagonal configurations also admit {\it mode channels\/}, which are
easily constructed from (\ref{6.27}) thanks to the orthonormality property
of the eigenvectors (\ref{6.24.b}):

\begin{equation}
  \hat y^{(m)}(s)\equiv\sum_{a=1}^5\,v_a^{(m)*}\hat q_a(s) =
  \eta^{-1/2}\,a_{n2}\,
  \frac{1}{2}\left[\left(s^2+\omega_{m+}^2\right)^{-1} -
  \left(s^2+\omega_{m-}^2\right)^{-1}\right]\,\hat g^{(2m)}(s) + O(0)
  \label{6.28}
\end{equation}

These are almost identical to the {\sl TIGA\/} mode channels (\ref{6.22}),
the only difference being that each mode channel comes now at a {\it single
specific\/} frequency pair $\omega_{m\pm}$. It thus appears that a
pentagonal transducer configuration enables signal observations over
a somewhat richer frequency band than does the {\sl TIGA\/}.

\begin{figure}[hbt]
\psfig{file=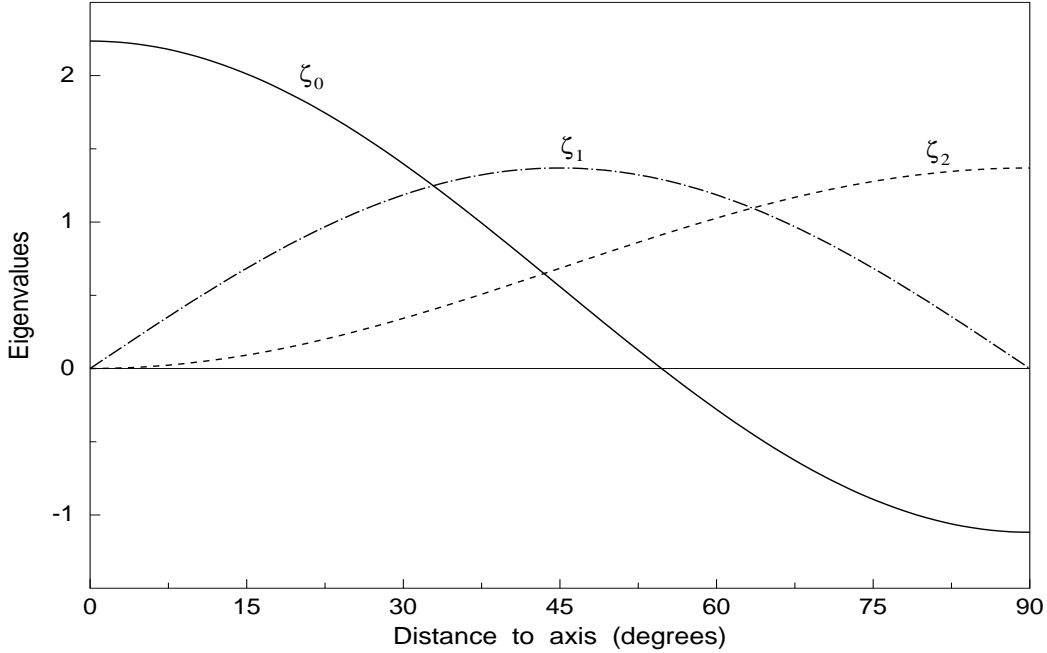,height=15cm,width=12cm,rheight=9.2cm,bbllx=-1.8cm,bblly=-3.8cm,bburx=18.4cm,bbury=24.7cm}
\caption{The three distinct eigenvalues $\zeta_m\/$ ($m\/$\,=\,0,1,2) as
functions of the distance of the resonator parallel's co-latitude $\alpha\/$
relative to the axis of symmetry of the distribution, cf. equation
(\protect\ref{6.24.a}).
\label{fig3}}
\end{figure}
\begin{figure}[htb]
\psfig{file=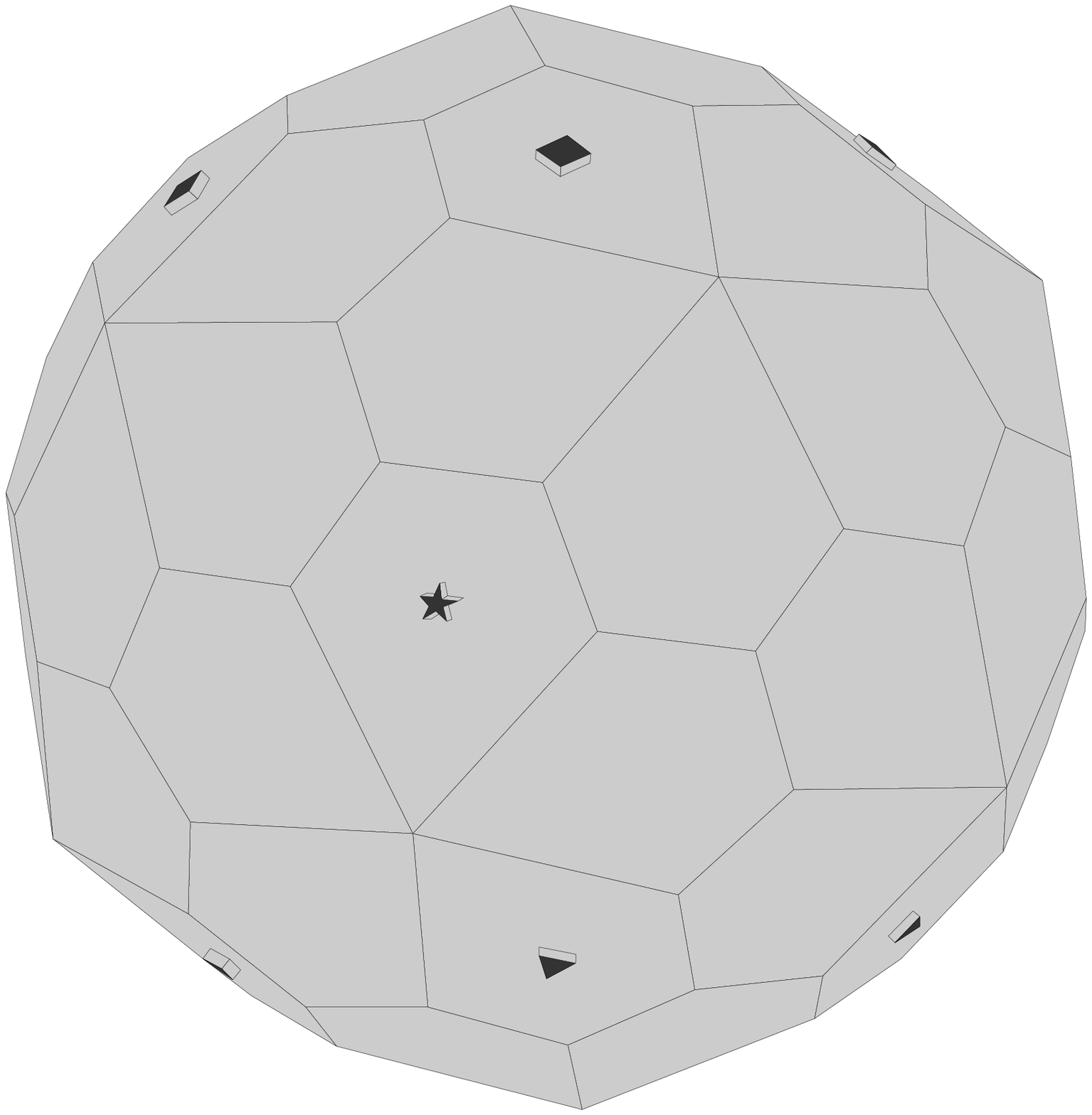,height=11cm,width=7cm}
\vspace*{-11 cm}
\hspace*{9.7 cm}
\psfig{file=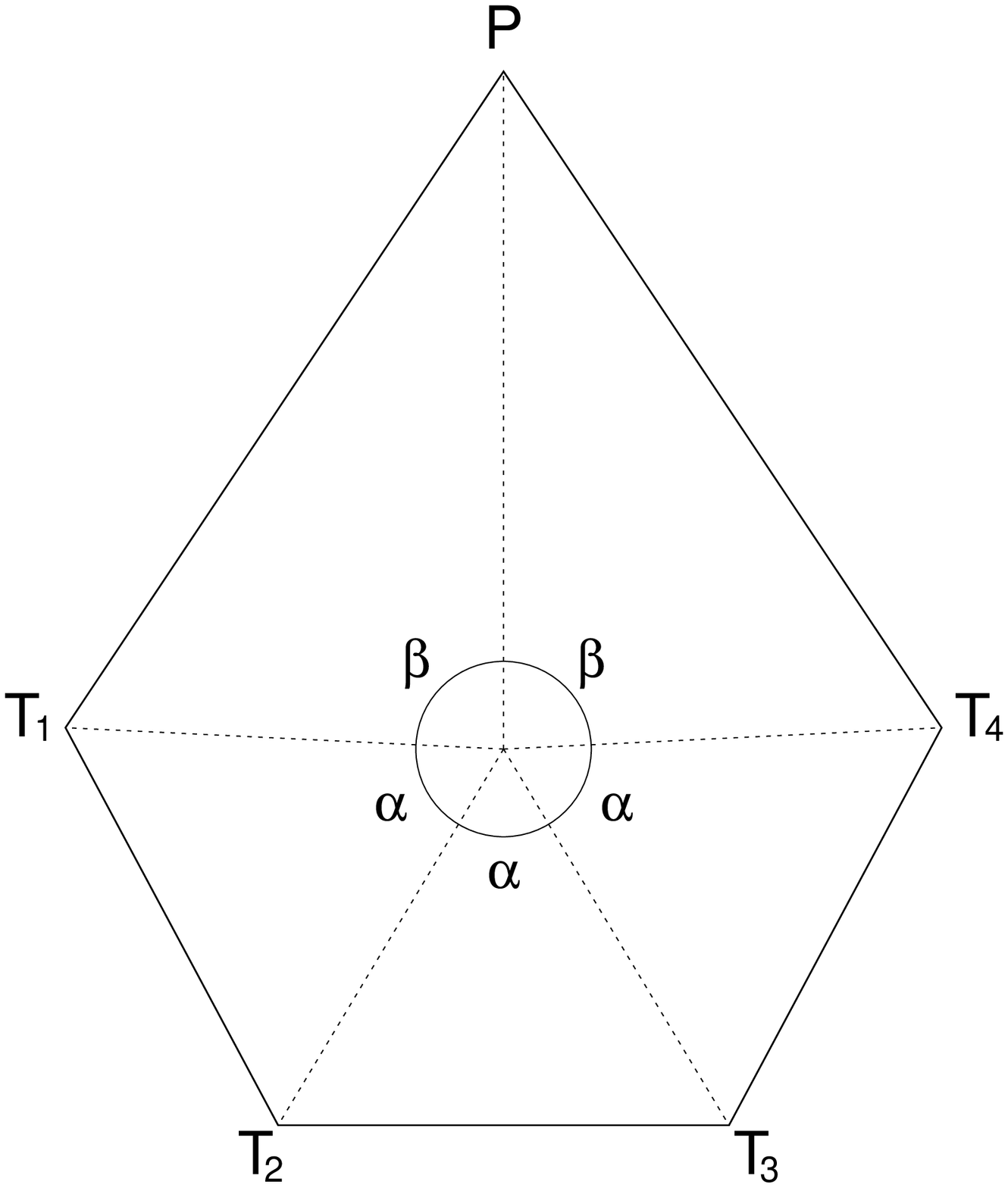,height=11cm,width=7cm}
\vspace*{-1.2 cm}
\caption{To the left, the {\it pentagonal hexacontahedron\/} shape. Certain
faces are marked to indicate resonator positions in a specific proposal
---see text--- as follows: a {\it square\/} for resonators tuned to the
first quadrupole frequency, a {\it triangle\/} for the second, and a
{\it star\/} for the monopole. On the right we see the (penatgonal) face
of the polyhedron. A few details about it: the confluence point of the
dotted lines at the centre is the tangency point of the {\it inscribed\/}
sphere to the {\sl PHC\/}; the labeled angles have values
$\alpha\/$\,=\,61.863$^\circ$, $\beta\/$\,=\,87.205$^\circ$; the angles at
the $T\/$-vertices are all equal, and their value is 118.1366$^\circ$,
while the angle at $P\/$ is 67.4536$^\circ$; the ratio of a long edge
(e.g. $PT_1$) to a short one (e.g. $T_1T_2$) is 1.74985, and the radius of
the inscribed sphere is {\it twice\/} the long edge of the pentagon,
$R\/$\,=\,2\,$PT_1$.	\label{fig4}}
\end{figure}
\begin{figure}[hbt]
\psfig{file=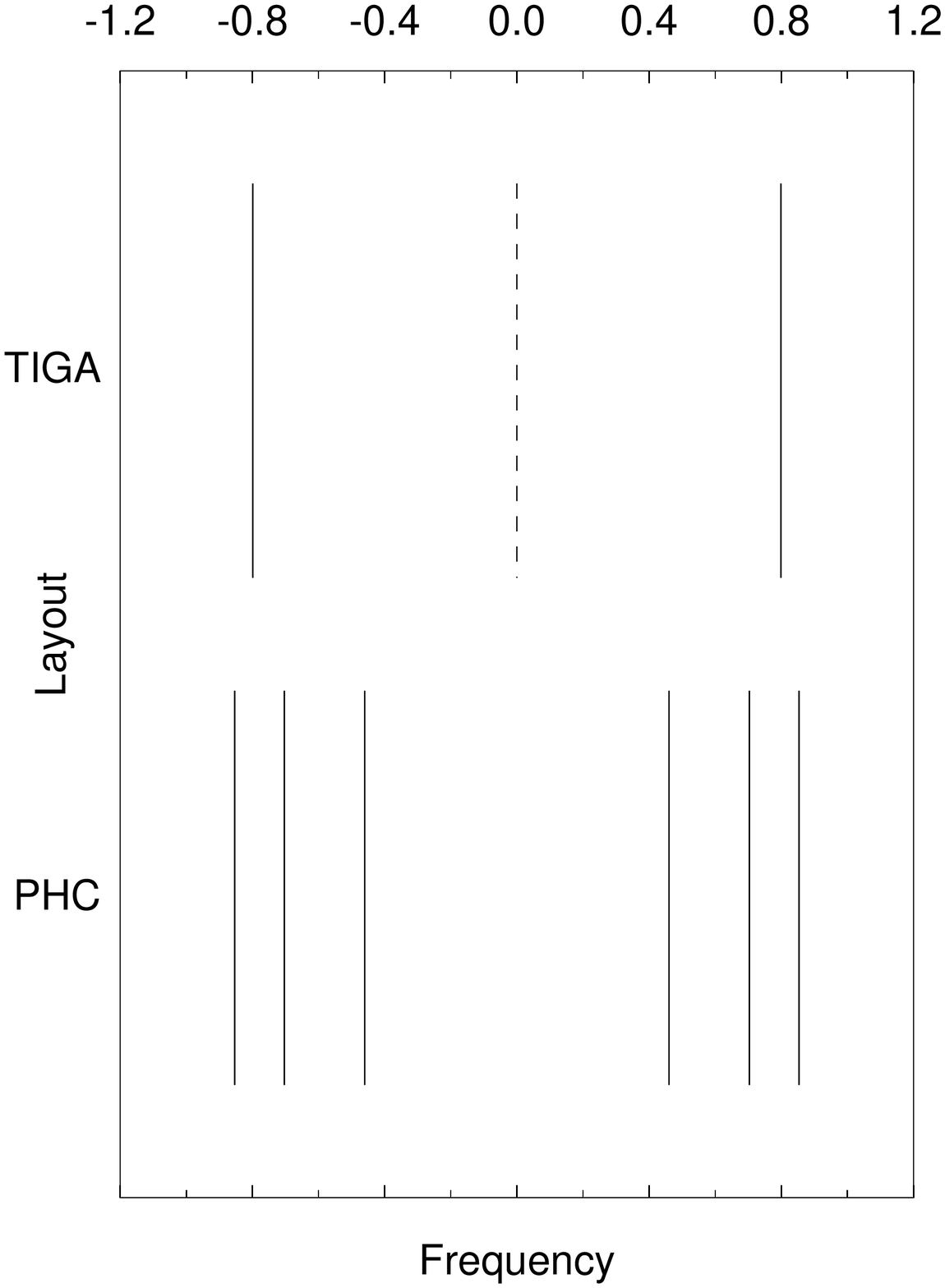,height=12cm,width=15cm,rheight=9.7cm,bbllx=-1cm,bblly=-5cm,bburx=20cm,bbury=25.8cm}
\caption{Compared line spectrum of a coupled {\sl TIGA\/} and a {\sl PHC\/}
resonator layout in an ideally spherical system. The weakly coupled central
frequency in the {\sl TIGA\/} is drawn dashed. The frequency pair is 5-fold
degenerate for this layout, while the two outer pairs of the {\sl PHC\/}
are doubly degenerate each, and the inner pair is non-degenerate. Units in
abscissas are $\eta^{1/2}\Omega$, and the central value, labeled 0.0,
corresponds to $\Omega$.
\label{fig5}}
\end{figure}

Based on these facts one may next ask which is a suitable transducer
distribution with an axis of pentagonal symmetry. In Figure \ref{fig3} we
give a plot of the eigenvalues (\ref{6.24.a}) as a function of $\alpha\/$,
the angular distance of the resonator set from the symmetry axis. Several
criteria may be adopted to select a specific choice in view of this graph.
An interesting one was proposed by us in reference \cite{ls} with the
following argument. If for ease of mounting, stability, etc., it is
desirable to have the detector milled into a close-to-spherical
{\it polyhedric\/} shape\footnote{
This is the philosophy suggested and experimentally implemented by
Merkowitz and Johnson at {\sl LSU\/}.}
then polyhedra with axes of pentagonal symmetry must be searched. The
number of quasi regular {\it convex\/} polyhedra is of course finite
---there actually are only 18 of them \cite{pacoM,tsvi}---, and we found
a particularly appealing one in the so called {\it pentagonal
hexacontahedron\/} ({\sl PHC\/}), which we see in Figure \ref{fig4}, left.
This is a 60 face polyhedron, whose faces are the identical {\it irregular
pentagons\/} on the right of the Figure. The {\sl PHC\/} admits an
{\it inscribed sphere\/} which is tangent to each face at the central
point marked in the Figure. It is, clearly, to this point that a resonator
should be attached so as to simulate an as perfect as possible spherical
distribution.

The {\sl PHC\/} is considerably spherical: the ratio of its volume to that
of the inscribed sphere is 1.057, which quite favourably compares to the
value of 1.153 for the ratio of the {\it circumscribed\/} sphere to the TI
volume. If we now request that the frequency pairs $\omega_{m\pm}$ be as
{\it evenly spaced\/} as possible, compatible with the {\sl PHC\/} face
orientations, then we must choose $\alpha\/$\,=\,67.617$^\circ$, whence

\begin{equation}
 \omega_{0\pm} = \omega_{12}\,\left(1\pm 0.5756\,\eta^{1/2}\right)\ ,\ \ \ 
 \omega_{1\pm} = \omega_{12}\,\left(1\pm 0.8787\,\eta^{1/2}\right)\ ,\ \ \ 
 \omega_{2\pm} = \omega_{12}\,\left(1\pm 1.0668\,\eta^{1/2}\right)
 \label{6.29}
\end{equation}

for instance for $\Omega$\,=\,$\omega_{12}$, the first quadrupole harmonic.
In Figure \ref{fig5} we display this frequency spectrum together with the
multiply degenerate {\it TIGA\/} for comparison. The frequency span of
both distributions is naturally comparable, yet the {\sl PHC\/} is
slightly broader.

The criterion leading to the {\sl PHC\/} proposal is of course not unique,
and alternatives can be considered. For example, if the 5 faces of a
regular icosahedron are selected for sensor mounting
($\alpha\/$\,=\,63.45$^\circ$) then a four-fold degenerate pair plus a
single non-degenerate pair is obtained; if the resonator parallel is
50$^\circ$ or 22.6$^\circ$ away from the ``north pole'' then the three
frequencies $\omega_{0+}$, $\omega_{1+}$, and $\omega_{2+}$ are equally
spaced; etc. The number of choices is virtually infinite if the sphere is
not milled into a polyhedric shape.

Let us finally recall that the complete {\sl PHC\/} proposal \cite{ls} was
made with the idea of building an as complete as possible spherical GW
antenna, which amounts to making it sensitive at the first {\it two\/}
quadrupole frequencies {\it and\/} at the first monopole one. This would
take advantage of the good sphere cross section at the second quadrupole
harmonic \cite{clo}, and would enable measuring (or thresholding)
eventual monopole GW radiation. Now, the system {\it pattern matrix\/}
$\hat\Lambda_a^{(lm)}(s;\Omega)$ has {\it identical structure\/} for all the
harmonics of a given $l\/$ series ---see (\ref{6.12}) and (\ref{6.15})---,
and so too identical criteria for resonator layout design apply to either
set of transducers, respectively tuned to $\omega_{12}$ and $\omega_{22}$.
The {\sl PHC\/} proposal is best described graphically in Figure \ref{fig4}:
a {\it second\/} set of resonators, tuned to the second quadrupole harmonic
$\omega_{22}$ can be placed in an equivalent position in the ``southern
hemisphere'', and an eleventh resonator tuned to the first monopole
frequency $\omega_{10}$ is added at an arbitrary position. It is not
difficult to see, by the general methods outlined earlier on in this paper,
that cross interaction between these three sets of resonators is only
{\it second order\/} in $\eta^{1/2}\/$, therefore weak.

A spherical GW detector with such a set of altogether 11 transducers would
be a very complete multi-mode multi-frequency device with an unprecedented
capacity as an individual antenna. Amongst other it would practically enable
monitoring of coalescing binary {\it chirp\/} signals by means of a rather
robust double passage method \cite{cf}, a prospect which was considered
so far possible only with broadband long baseline laser interferometers
\cite{klm1,klm2}, and is almost unthinkable with currently operating
cylindrical bars.

\section{A calibration signal: hammer stroke}

This section is a brief digression from the main streamline of the paper.
We propose to assess now the system response to a particular, but useful,
calibration signal: a perpendicular {\it hammer stroke\/}. After a few
general considerations we describe the situation in a {\sl TIGA\/} and in
a {\sl PHC\/} configuration.

We first go back to equation (\ref{3.16}) and replace
$\hat u_a^{\rm external}(s)$ in its rhs with that corresponding to a
hammer stroke, which is easily calculated thanks to (\ref{3.4.c}),
(\ref{3.5.c}), and (\ref{3.165}); the result is

\begin{equation}
  \hat u_a^{\rm stroke}(s) = -\sum_{nl}\,\frac{f_0}{s^2+\omega_{nl}^2}\,
  \left|A_{nl}(R)\right|^2\,P_l({\bf n}_a\!\cdot\!{\bf n}_0)\ ,\qquad
  a=1,\ldots,J  \label{7.1}
\end{equation}
where ${\bf n}_0$ are the spherical coordinates of the hit point on the
sphere, and $f_0$\,$\equiv$\,${\bf n}_0\!\cdot\!{\bf f}_0/{\cal M\/}$.
Clearly, the hammer stroke excites {\it all\/} of the sphere's vibration
eigenmodes, as it has a completly flat spectrum.

The coupled system resonances are of course those already calculated in
section 6, and the same procedures described there for a GW excitation
can be pursued now to obtain

\begin{eqnarray}
  \hat q_a(s) & = & \eta^{-1/2}\,(-1)^{J-1}\,\sqrt{\frac{2l+1}{4\pi}}
  \,f_0\,\left|A_{nl}(R)\right|\,\times  \nonumber \\
  & \times & \sum_{b=1}^J\,\left\{\sum_{\zeta_c\neq 0}\,\frac{1}{2}\left[
  \left(s^2+\omega_{c+}^2\right)^{-1}-\left(s^2+\omega_{c-}^2\right)^{-1}
  \right]\,\frac{v_a^{(c)}v_b^{(c)*}}{\zeta_c}\right\}\,
  P_l({\bf n}_b\!\cdot\!{\bf n}_0) + O(0)\ ,
  \qquad a=1\,\ldots,J	\label{7.2}
\end{eqnarray}

when the system is tuned to the $nl\/$-th spheroidal harmonic, i.e.,
$\Omega$\,=\,$\omega_{nl}$. It is immediately seen from here that the
system response to this signal when the resonators are tuned to a
monopole frequency is given by

\begin{equation}
  \hat q_a(s) = \eta^{-1/2}\,(-1)^{J-1}\,\frac{f_0}{\sqrt{4\pi J}}\,
  \left|A_{n0}(R)\right|\,\frac{1}{2}\left[\left(
  s^2+\omega_+^2\right)^{-1}-\left(s^2+\omega_-^2\right)^{-1}\right]
  \ ,\qquad \Omega=\omega_{n0}
  \label{7.3}
\end{equation}
an expression which holds for all $a\/$, and is independent of either the
resonator layout or the hit point, which in particular prevents any
determination of the latter, as obviously expected. The frequencies
$\omega_\pm$ are those of (\ref{6.11}), and we find here again a global
factor $J^{-1/2}$, as also expected.

We consider next the situation when quadrupole tuning is implemented,
$\Omega$\,=\,$\omega_{n2}$. We shall however do so only for the {\sl TIGA\/}
and {\sl PHC\/} configurations, as more general considerations are not
quite as interesting at this point.

\subsection{{\sl TIGA\/} response to a hammer stroke}

One easily verifies that, for the TI configuration,

\begin{equation}
  \hat q_a(s) = -\eta^{-1/2}\,\frac{5}{\sqrt{24\pi}}\,
  f_0\,\left|A_{n2}(R)\right|\,\frac{1}{2}\left[
  \left(s^2+\omega_+^2\right)^{-1}-\left(s^2+\omega_-^2\right)^{-1}\right]
  \,P_2({\bf n}_a\!\cdot\!{\bf n}_0)\ ,\qquad\text{\sl TIGA}
  \label{7.4}
\end{equation}
where $\omega_\pm$ are the {\sl TIGA\/} frequencies (\ref{6.19}), and
$\Omega$\,=\,$\omega_{n2}$. This equation shows that the hitting point
position ${\bf n}_0$ {\it can\/} be easily determined from the readouts
$\hat q_a(s)$ ---as a matter of fact it is {\it redundantly\/} determined
by them. The {\it mode channels\/} for this signal yield

\begin{equation}
  \hat y^{(m)}(s) = -\eta^{-1/2}\,f_0\,\left|A_{n2}(R)\right|
  \frac{1}{2}\left[\left(s^2+\omega_+^2\right)^{-1} -
  \left(s^2+\omega_-^2\right)^{-1}\right]\,Y_{2m}^*({\bf n}_0)
  \ ,\ \  m=-2,\ldots,2	\label{7.5}
\end{equation}
and they are proportional to the sphere's quadrupole radial oscillation
amplitude $A_{n2}(R)\,Y_{2m}^*({\bf n}_0)$ at the hit point, ${\bf n}_0$
\cite{lobo}. A simple numerical simulation is illustrative of the situation,
and we present it here.

\begin{figure}[htb]
\psfig{file=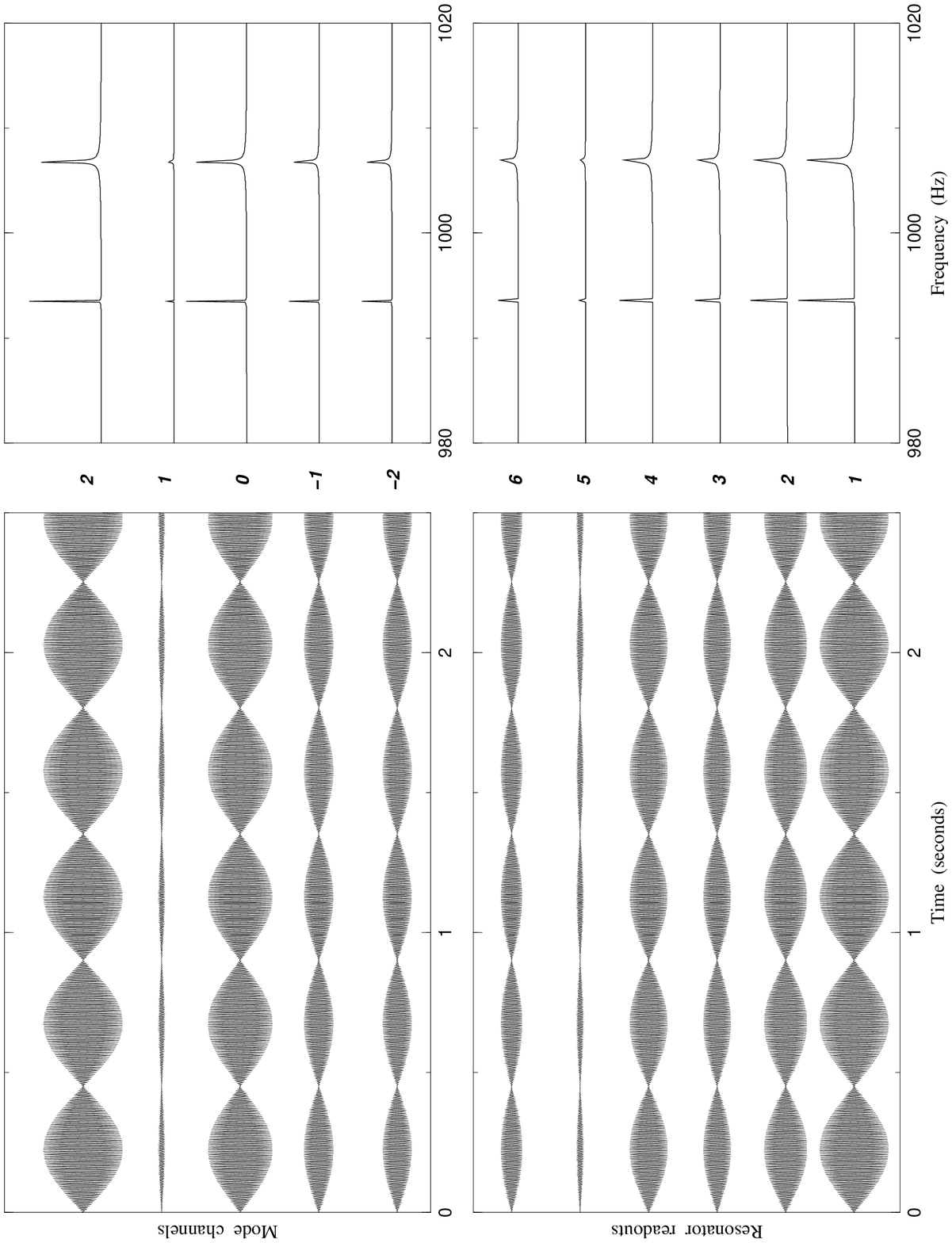,height=22cm,width=15cm,rheight=23cm,bbllx=1.1cm,bblly=3.3cm,bburx=17.3cm,bbury=27cm}
\caption{Simulated response of a {\sl TIGA\/} to a hammer stroke: the time
series and their respective spectra, both for direct resonator readouts and
mode channels. Asymmetric widening of spectral lines is due to frequency
{\it leakage\/} caused by finite integration time.
\label{fig6}}
\end{figure}

We recall that the inverse Laplace transform of $2^{-1}$$\left[\left(
s^2+\omega_+^2\right)^{-1}-\left(s^2+\omega_-^2\right)^{-1}\right]$ is

\begin{equation}
  \Omega^{-1}\,\sin\frac{1}{2}(\omega_+-\omega_-)t\,\cos\Omega t
  + O(\eta^{1/2})       \label{7.6}
\end{equation}

which is a {\it beat\/} ---a sinusoid of carrier frequency $\Omega$
amplitude modulated by another sinusoid of much smaller frequency of order
$\eta^{1/2}\Omega$. All six resonators' motions are thus identical beats,
except that amplitude varies from one to the other, and the same applies
to all five mode channels. We see this graphically in Figure \ref{fig6},
wherein plots are displayed of the resonator readouts and mode channels.

\subsection{{\sl PHC\/} response to a hammer stroke}

The situation is slightly more involved for a {\sl PHC\/} distribution,
but still easy to address; the system response is given by

\begin{equation}
  \hat q_a(s) = \eta^{-1/2}\,f_0\,\sqrt{\frac{4\pi}{5}}\,
  \left|A_{n2}(R)\right|\,\sum_{m=-2}^2\,\frac{1}{2}\left[\left(
  s^2+\omega_{m+}^2\right)^{-1}-\left(s^2+\omega_{m-}^2\right)^{-1}\right]
  \,\zeta_m^{-1}\,Y_{2m}({\bf n}_a)\,Y_{2m}^*({\bf n}_0)
  \ ,\qquad\text{\sl PHC}
  \label{7.7}
\end{equation}

and the mode channels by

\begin{equation}
  \hat y^{(m)}(s) = \eta^{-1/2}\,f_0\,\left|A_{n2}(R)\right|
  \frac{1}{2}\left[\left(s^2+\omega_{m+}^2\right)^{-1} -
  \left(s^2+\omega_{m-}^2\right)^{-1}\right]\,Y_{2m}^*({\bf n}_0)
  \ ,\ \ m=-2,\ldots,2	\label{7.8}
\end{equation}

\begin{figure}[htb]
\psfig{file=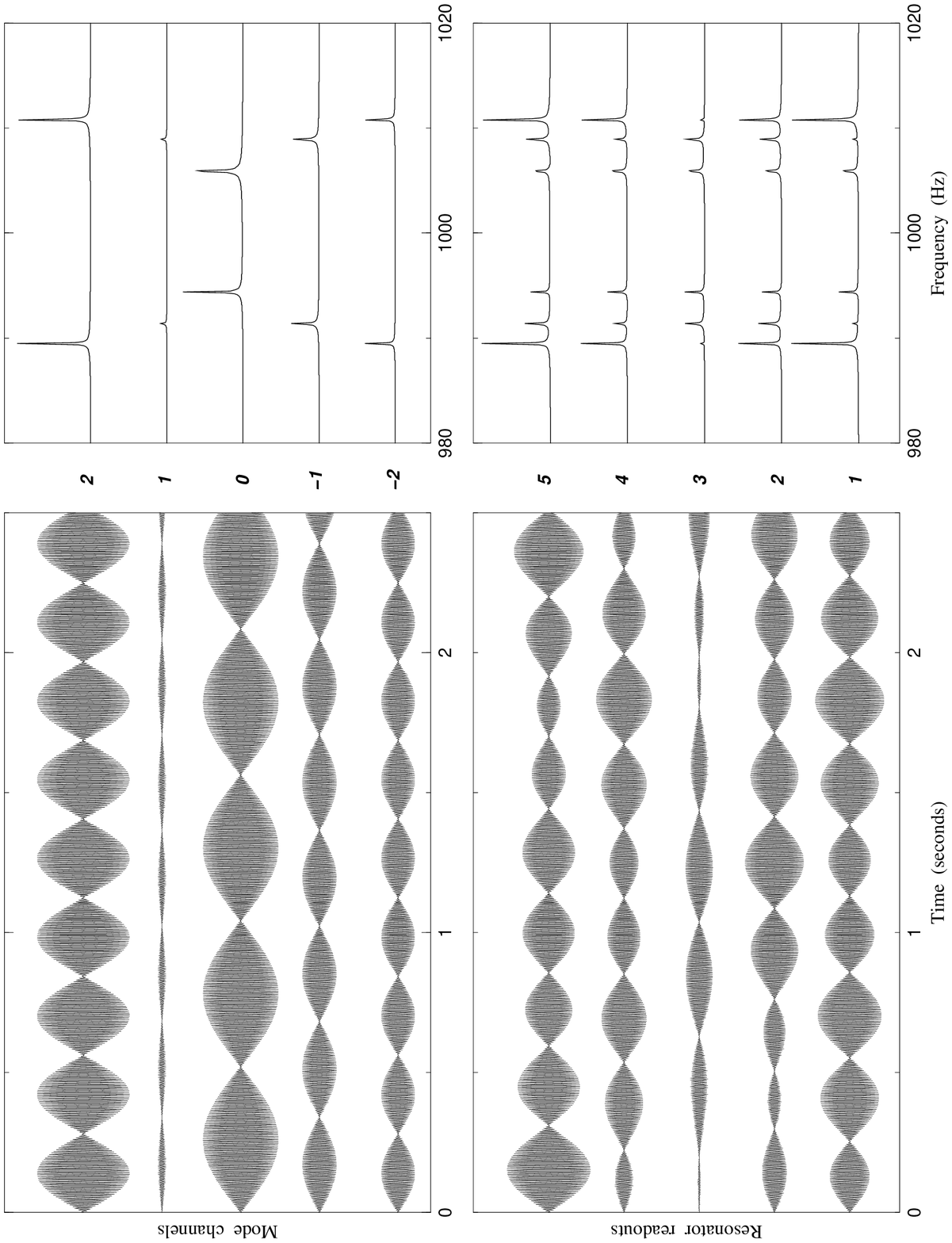,height=22cm,width=15cm,rheight=23cm,bbllx=1.1cm,bblly=3.3cm,bburx=17.3cm,bbury=27cm}
\caption{Simulated response of a {\sl PHC\/} to a hammer stroke: the time
series and their respective spectra, both for direct resonator readouts and
mode channels. Note that while the former are {\it not\/} simple beats, the
latter are.	\label{fig7}}
\end{figure}

The difference with the {\sl TIGA\/} is this: the system response $q_a(t)$
is a {\it superposition of three different beats\/}, while the mode
channels are {\it single\/} beats each, but with {\it differing modulation
frequencies\/}. This is represented graphically in Figure \ref{fig7},
where we see the result of a numerical simulation of the {\sl PHC\/}
response to a hammer stroke, delivered to the solid at the same location
as in the {\sl TIGA\/} example of the previous subsection. The readouts
$q_a(t)$ are somewhat fancy time series, whose frequency spectrum shows
{\it three pairs of peaks\/} ---in fact, the {\it lines\/} in the ideal
spectrum of Figure \ref{fig5}. The mode channels on the other hand are
{\it pure beats\/}, whose spectra consist of the {\it individually
separate\/} pairs of previous peaks. It might perhaps be said that the
{\sl PHC\/} gives rise to a sort of ``Zeeman splitting'' of the {\sl TIGA\/}
degenerate frequencies, which can be attributed to an {\it axial symmetry
breaking\/} of the isotropic character of that resonator distribution; the
{\sl PHC\/} mode channels naturally resolve the split multiplet into its
components.

This simple example very neatly displays the most relevant features of
both {\sl TIGA\/} and {\sl PHC\/} resonator distributions, and enables a
quite clear comparison of the merits of either, which does apply also to
real GW signals. Despite their distinct characteristics, these layouts
are perfectly equivalent as regards their ability to sense GWs {\it in an
ideally noiseless\/} detector. This is so simply because the detector is
the sphere, not the transducers, and therefore {\it any\/} motion sensing
system, not just these two, is equivalent to them if perfectly noiseless
monitoring were possible. Things do however change in the presence of
noise. T.\ Stevenson has recently addressed the problem of how isotropy in
sensitivity is affected by noise \cite{ts}, and concludes that while the
{\sl TIGA\/} maintains isotropic sensitivity, the {\sl PHC\/} does not: the
latter is slightly more sensitive in a relatively small solid angle around
the resonators' axis, but slightly less for other incidence directions.
This is a clear disadvantage of the {\sl PHC\/}, but one may not forget
that a real antenna will {\it not\/} be spherically symmetric because it
must be {\it suspended\/}, and that breaks that symmetry. As we shall
shortly see, the {\sl PHC\/} is very naturally adapted to that symmetry
breaking, and therefore it is probably unwise to outrightly dispose of it
on the basis of a single noise criterion. At the same time, the {\it PHC\/}
system spectrum is richer, and slightly broader than the {\sl TIGA\/}'s,
and this again favours {\sl PHC\/}, even in the presence of noise, as the
possibility of looking for signals at more frequencies reduces the
probability of random disturbances, thereby increasing the probability of
signal detection.

The above is a {\it qualitative\/} discussion, whose {\it quantitative\/}
aspects will not be addressed in any more depth in this paper. We still
wish to assess the effects of imperfections in a noiseless detector,
which are of paramount relevance in a real system.

\section{Symmetry deffects}

So far we have made the assumption that the sphere is perfectly symmetric,
that the resonators are identical, that their locations on the sphere's
surface are ideally accurate, etc. This is of course unrealistic. So we
propose to address now how such departures from ideality affect the
system behaviour. As we shall see, the system is rather {\it robust\/},
in a sense to be made precise shortly, against a number of small deffects.

In order to {\it quantitatively\/} assess ideality failures we shall adopt
a philosophy which is naturally suggested by the results already obtained
in an ideal system. It is as follows.

As we have seen in previous sections, the solution to the general equations
(\ref{3.16}) must be given as a {\it perturbative\/} series expansion in
ascending powers of the small quantity $\eta^{1/2}$. This is clearly not a
fact related to the system's symmetries, and so it will survive symmetry
breakings. It is therefore appropriate to {\it parametrise\/} deviations
from ideality in terms of suitable powers of $\eta^{1/2}$, in order to
address them {\it consistently with the order of accuracy of the series
solution to the equations of motion\/}. An example will better illustrate
the situation.

In a {\it perfectly ideal\/} spherical detector the system frequencies
are given by equations (\ref{5.2}). Now, if a small departure from e.g.
spherical symmetry is present in the system then we expect that a
correspondingly small correction to those equations will be required.
Which specific correction to the formula will actually happen can be
{\it qualitatively\/} assessed by a {\it consistency\/} argument: if
symmetry deffects are of order $\eta^{1/2}$ then equations (\ref{5.2}) will
be significantly altered in their $\eta^{1/2}$ terms; if on the other hand
such deffects are of order $\eta\/$ or smaller then any modifications to
equations (\ref{5.2}) will be swallowed into the $O(0)$ terms, and the
more important $\eta^{1/2}$ terms will remain unaffected by this symmetry
failure. We will say in the latter case that the system is {\it robust\/}
to that ideality breaking.

More generally, this argument can be extended to see that the only system
deffects standing a chance to have any influences on lowest order ideal
system behaviour are deffects of order $\eta^{1/2}$ relative to an ideal
configuration. Deffects of such order are however {\it not necessarily
guaranteed\/} to be significant, and a specific analysis is required for
each specific parameter in order to see whether or not the system response
is {\it robust\/} against the considered parameter deviations.

We therefore proceed as follows. Let $P\/$ be one of the system parameters,
e.g. a sphere frequency, or a resonator mass or location, etc. Let
$P_{\rm ideal}$ be the {\it numerical value\/} this parameter has in an
ideal detector, and let $P_{\rm real}$ be its value in the real case. These
two will be assumed to differ by terms of order $\eta^{1/2}$, or

\begin{equation}
  P_{\rm real} = P_{\rm ideal}\,(1+p\,\eta^{1/2})     \label{8.1}
\end{equation}

For a given system, $p\/$ is readily determined adopting (\ref{8.1}) as the
{\it definition\/} of $P_{\rm real}$, once a suitable {\it hypothesis\/}
has been made as to which is the value of $P_{\rm ideal}$. In order for
the following procedure to make sensible sense it is clearly required that
$p\/$ be of order 1 or, at least, appreciably larger than $\eta^{1/2}$.
Should $p\/$ thus calculated from (\ref{8.1}) happen to be too small, i.e.,
of order $\eta^{1/2}$ itself or smaller, then the system will be considered
{\it robust\/} as regards the affected parameter.

We now apply this criterion to various departures from ideality.

\subsection{The suspended sphere  \label{s8.1}}

An earth based observatory obviously requires a {\it suspension
mechanism\/} for the large sphere. If a {\it nodal point\/} suspension
is e.g.\ selected then a diametral {\it bore\/} has to be drilled across
the sphere \cite{phd}. The most immediate consequence of this is that
spherical symmetry is broken, what in turn results in {\it degeneracy
lifting\/} of the free spectral frequencies $\omega_{nl}\/$, which now
{\it split\/} up into multiplets $\omega_{nlm}\/$
($m\/$\,=\,$-l\/$,...,$l\/$). The resonators' frequency $\Omega$
{\it cannot\/} therefore be matched to {\it the\/} frequency
$\omega_{n_0l_0}$, but at most to {\it one\/} of the
$\omega_{n_0l_0m}\/$'s. In this subsection we keep the hypohesis that
all the resonators are identical ---we shall relax it later---, and
assume that $\Omega$ falls {\it within\/} the span of the multiplet
of the $\omega_{n_0l_0m}\/$'s. Then we write

\begin{equation}
  \omega_{n_0l_0m}^2 = \Omega^2\,(1+p_m\,\eta^{1/2})\ ,\qquad
  m=-l_0,\ldots,l_0     \label{8.2}
\end{equation}

We now search for the coupled frequencies, i.e., the roots of equation
(\ref{3.18}). The kernel matrix $\hat K_{ab}(s)$ is however no longer
given by (\ref{4.2}), due the removed degeneracy of $\omega_{nl}\/$, and
we must stick to its general expression (\ref{3.17}), or

\begin{equation}
  \hat K_{ab}(s) = \sum_{nlm}\,\frac{\Omega_b^2}{s^2+\omega_{nlm}^2}\,
   \left|A_{nl}(R)\right|^2\,\frac{2l+1}{4\pi}\,
   Y_{lm}^*({\bf n}_a)\,Y_{lm}({\bf n}_b) \equiv
   \sum_{nlm}\,\frac{\Omega_b^2}{s^2+\omega_{nlm}^2}\,\chi_{ab}^{(nlm)}
   \label{8.3}
\end{equation}

Following the steps of section 4 we now need to seek the roots of the
equation

\begin{equation}
  \det\,\left[\delta_{ab} + \eta\,\sum_{m=-l_0}^{l_0}\,
   \frac{\Omega^2s^2}{(s^2+\Omega^2)(s^2+\omega_{n_0l_0m}^2)}
   \,\chi_{ab}^{(n_0l_0m)} + \eta\,\sum_{nl\neq n_0l_0,m}\,
   \frac{\Omega^2s^2}{(s^2+\Omega^2)(s^2+\omega_{nlm}^2)}\,
   \chi_{ab}^{(nlm)}\right] = 0
  \label{8.4}
\end{equation}

Since $\Omega$ relates to $\omega_{n_0l_0m}\/$ through equation (\ref{8.2})
we see that the roots of (\ref{8.4}) follow again into either of the two
categories (\ref{4.11}), i.e., roots close to $\pm i\Omega$ and roots close
to $\pm i\omega_{nlm}\/$ ($nl\/$\,$\neq$\,$n_0l_0$). We shall exclusively
concentrate on the former now. Direct substitution of the series
(\ref{4.11.a}) into (\ref{8.4}) yields the following equation for the
coefficient $\chi_{\frac{1}{2}}$:

\begin{equation}
  \det\left[\delta_{ab} - \frac{1}{\chi_\frac{1}{2}}\,\sum_{m=-l_0}^{l_0}
  \,\frac{\chi_{ab}^{(n_0l_0m)}}{\chi_\frac{1}{2}-p_m}\right] = 0
  \label{8.5}
\end{equation}

This is a variation of (\ref{5.1}), to which it reduces when
$p_m\/$\,=\,0, i.e., when there is full degeneracy.

The solutions to (\ref{8.5}) no longer come in symmetric pairs, like
(\ref{5.2}). Rather, there are 2$l_0$+1+$J\/$ of them, with a
{\it maximum\/} number of 2(2$l_0$+1) non-identically zero roots if
$J\/$\,$\geq$\,2$l_0$+1\footnote{
This is a {\it mathematical fact\/}, whose proof is relatively cumbersome,
and will be omitted here; we just mention that it has its origin in the
linear dependence of more than 2$l_0$+1 spherical harmonics of order $l_0$.}.
For example, if we choose to select the resonators' frequency close to a
quadrupole multiplet ($l_0$\,=\,2) then (\ref{8.5}) has at most 5+$J\/$
non-null roots, {\it with a maximum ten\/} no matter how many resonators
in excess of 5 we attach to the sphere. Modes associated to null roots of
(\ref{8.5}) can be seen to be {\it weakly coupled\/}, just like in a free
sphere, i.e., their amplitudes are smaller than those of the strongly
coupled ones by factors of order $\eta^{1/2}$.

In order to assess the reliability of this method we have applied it to
see what are its predictions for a {\it real system\/}. To this end, data
taken with the {\sl TIGA\/} prototype at {\sl LSU\/}\footnote{
These data are contained in reference \protect\cite{phd}, and we want to
express our gratitude to Stephen Merkowitz for kindly handing them to us.}
were used to confront with. The {\sl TIGA\/} was drilled and suspended
from its centre, so its first quadrupole frequency split up into a
multiplet of five frequencies. Their reportedly measured values are

\begin{equation}
  \omega_{120} = 3249\ {\rm Hz}\ ,\ \ 
  \omega_{121} = 3238\ {\rm Hz}\ ,\ \ 
  \omega_{12\,-1} = 3236\ {\rm Hz}\ ,\ \ 
  \omega_{122} = 3224\ {\rm Hz}\ ,\ \ 
  \omega_{12\,-2} = 3223\ {\rm Hz}\ ,\ \ 
  \label{8.6}
\end{equation}

All 6 resonators were equal, and had the following characteristic
frequency and mass, respectively:

\begin{equation}
  \Omega = 3241\ {\rm Hz}\ ,\qquad\eta = \frac{1}{1762.45}
  \label{8.7}
\end{equation}

Substituting these values into (\ref{8.2}) it is seen that

\begin{equation}
  p_0=0.2075\ ,\ \   p_1=-0.0777\ ,\ \   p_{-1}=-0.1036\ ,\ \ 
  p_2=-0.4393\ ,\ \  p_{-2}=-0.4650
  \label{8.8}
\end{equation}

\begin{table}[tb]
\caption{Numerical values of measured and theoretically predicted
frequencies (in Hz) for the {\sl TIGA\/} prototype with varying number of
resonators. Percent differences are also shown. The {\it calculated\/}
values of the tuning and free multiplet frequencies are taken {\it by
definition\/} equal to the measured ones, and quoted in brackets. In
square brackets the frequency of the {\it weakly coupled\/} sixth mode
in the full, 6 resonator {\sl TIGA\/} layout. These data are plotted in
Figure \protect\ref{fig8}.    \label{t1}}

\begin{tabular}{lddd||lddd}
Item & Measured (Hz) & Calculated (Hz) & \% difference &
Item & Measured (Hz) & Calculated (Hz) & \% difference \\
\hline Tuning & 3241 & (3241) & (0.00) &
4 resonators & 3159 & 3155 & -0.12 \\
Free multiplet & 3223 & (3223)  & (0.00) &
             & 3160 & 3156 & -0.11 \\
               & 3224 & (3224)  & (0.00) &
             & 3168 & 3165 & -0.12 \\
               & 3236 & (3236)  & (0.00) &
             & 3199 & 3198 & -0.05 \\
               & 3238 & (3238)  & (0.00) &
             & 3236 & 3236 & 0.00 \\
               & 3249 & (3249)  & (0.00) &
             & 3285 & 3286 & 0.03 \\
1 resonator  & 3167 & 3164 & -0.08 &
             & 3310 & 3310 & 0.00 \\
             & 3223 & 3223 & 0.00 &
             & 3311 & 3311 & 0.00 \\
             & 3236 & 3235 & -0.02 &
             & 3319 & 3319 & 0.00 \\
             & 3238 & 3237 & -0.02 &
5 resonators & 3152 & 3154 & 0.08 \\
             & 3245 & 3245 & 0.00 &
             & 3160 & 3156 & -0.14 \\
             & 3305 & 3307 & 0.06 &
             & 3163 & 3162 & -0.03 \\
2 resonators & 3160 & 3156 & -0.13 &
             & 3169 & 3167 & -0.08 \\
             & 3177 & 3175 & -0.07 &
             & 3209 & 3208 & -0.02 \\
             & 3233 & 3233 & 0.00 &
             & 3268 & 3271 & 0.10 \\
             & 3236 & 3236 & 0.00 &
             & 3304 & 3310 & 0.17 \\
             & 3240 & 3240 & 0.00 &
             & 3310 & 3311 & 0.03 \\
             & 3302 & 3303 & 0.03 &
             & 3313 & 3316 & 0.10 \\
             & 3311 & 3311 & 0.00 &
             & 3319 & 3321 & 0.06 \\
3 resonators & 3160 & 3155 & -0.15 &
6 resonators & 3151 & 3154 & 0.11 \\
             & 3160 & 3156 & -0.13 &
             & 3156 & 3155 & -0.03 \\
             & 3191 & 3190 & -0.02 &
             & 3162 & 3162 & 0.00 \\
             & 3236 & 3235 & -0.02 &
             & 3167 & 3162 & -0.14 \\
             & 3236 & 3236 & 0.00 &
             & 3170 & 3168 & -0.07 \\
             & 3297 & 3299 & 0.08 &
             & [3239] & [3241] & [0.06] \\
             & 3310 & 3311 & 0.02 &
             & 3302 & 3309 & 0.23 \\
             & 3311 & 3311 & 0.00 &
             & 3308 & 3310 & 0.06 \\
             &  &  &  &
             & 3312 & 3316 & 0.12 \\
             &  &  &  &
             & 3316 & 3317 & 0.02 \\
             &  &  &  &
             & 3319 & 3322 & 0.10
\end{tabular}
\end{table}

\begin{figure}[htb]
\psfig{file=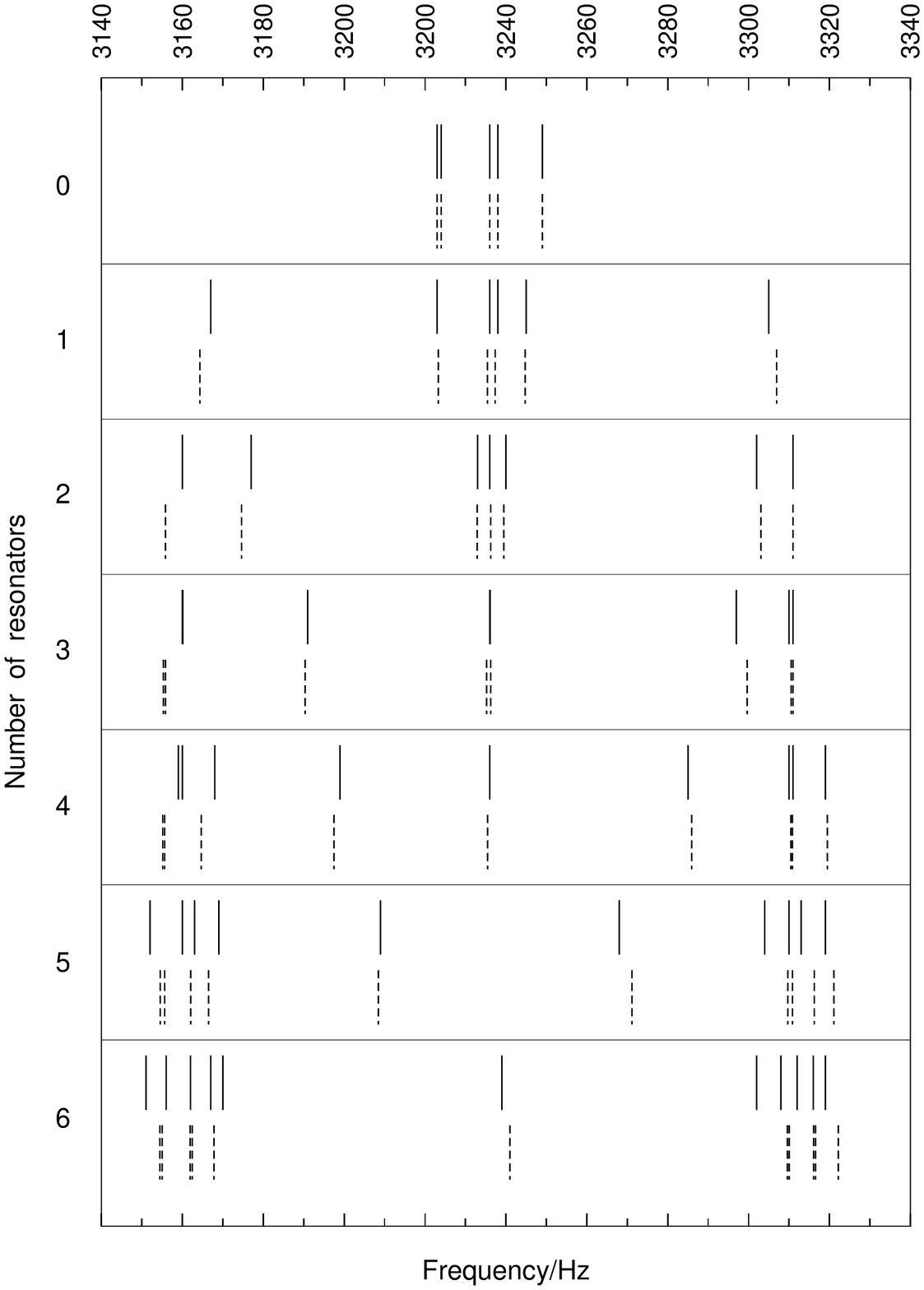,height=22cm,width=15cm,rheight=22cm}
\caption{The frequency spectrum of the {\sl TIGA\/} distribution as
resonators are progressively added from none to 6. Continuous lines
correspond to measured values, and dashed lines correspond to their
$\eta^{1/2}$ theoretical estimates of equation (\protect\ref{8.5}).
\label{fig8}}
\end{figure}

Equation (\ref{8.5}) can now be readily solved once the resonator
positions are fed into the matrices $\chi_{ab}^{(12m)}$. Such positions
correspond to the pentagonal faces of a truncated icosahedron. Merkowitz
\cite{phd} gives a complete account of all the measured system frequencies
as resonators are progressively attached to the selected faces, beginning
with one and ending with six. In Figure \ref{fig8} we present a graphical
display of the experimentally reported frequencies along with those
calculated theoretically by solving equation (\ref{8.5}). In Table 1 we
give the numerical values. As can be seen, coincidence between our
theoretical predictions and the experimental data is remarkable: the
worst error is 0.2\%, while for the most part it is below 0.1\%. This
should be taken as very strong evidence that our theoretical model
{\it is\/} correct, since {\it discrepancies between its predictions and
experiment are of order $\eta$\/}, as indeed expected ---see (\ref{8.7}).
In addition, it is also reported in reference \cite{jm97} that the 11-th,
weakly coupled mode of the {\sl TIGA\/} (highlighted in square brackets in
Table \ref{t1}) has a practically zero ampliutde, again in excellent
agreement with our general theoretical predictions about modes beyond the
tenth ---see paragraph after equation (\ref{8.5}).

This is an encouraging result which motivated us to try a better fit by
estimating the {\it next order\/} corrections, i.e., $\chi_1$ of
(\ref{4.11.a}). As it turned out, however, matching between theory and
experiment did not consistently improve. This is not really that surprising,
though, as M\&J explicitly state \cite{jm97} that control of the general
experimental conditions in which data were obtained had a certain degree of
tolerance, and they actually show satisfaction that $\sim$1\% coincidence
between theory and measurement is comfortably accomplished. But 1\% is
{\it two orders of magnitude larger than $\eta\/$} ---cf. equation
(\ref{8.7})---, so failure to refine our frequency estimates to order
$\eta\/$ is fully consistent with the accuracy of available real data.

We do expect that the analytical procedures developed in this paper will
be the appropriate ones to assess and theoretically understand the system
behaviour as rigor in system parameter control is progressively gained.

A final word on a technical issue is in order. Merkowitz and Johnson's
equations for the {\sl TIGA\/} \cite{jm93,jm95} are identical to ours to
lowest order in $\eta\/$. Remarkably, though, their reported theoretical
estimates of the system frequencies are not quite as accurate as ours.
The reason for this is probably the following: in M\&J's model these
frequencies appear within an algebraic system of 5+$J\/$ linear equations
with as many unknowns which has to be solved; in our model the algebraic
system has only $J\/$ equations and unknowns, actually equations
(\ref{3.16}). This is a very appreciable difference indeed for the range
of values of $J\/$ under consideration. While the roots for the frequencies
can be seen to {\it mathematically\/} coincide in both approaches, in
actual practice these roots are {\it estimated\/}, generally by means of
computer programmes. It is here that problems most likely arise, for the
numerical reliability of an algorithm to solve matrix equations normally
decreases as the rank of the matrix increases. The significant algebraic
simplification of our model's equations should therefore be considered
one of real practical value.

\subsubsection{The suspended {\sl PHC}}

We briefly consider now which would be the effects of symmetry breaking
due to suspension in a {\sl PHC\/} resonator distribution. The natural
suspension axis is the symmetry axis of the resonators, so we shall
assume that 5 of them are attached to the detector around that axis as
described in the paragraphs before equation (\ref{6.29}). To be specific,
we shall speculate with a {\sl PHC\/} having the numerical parameters of
equations (\ref{8.6}) and (\ref{8.7}), a hypothetical but reasonable
conjecture which will enable comparison with the actual {\sl TIGA\/}
prototype.

\begin{figure}[htb]
\hspace*{1 cm}
\psfig{file=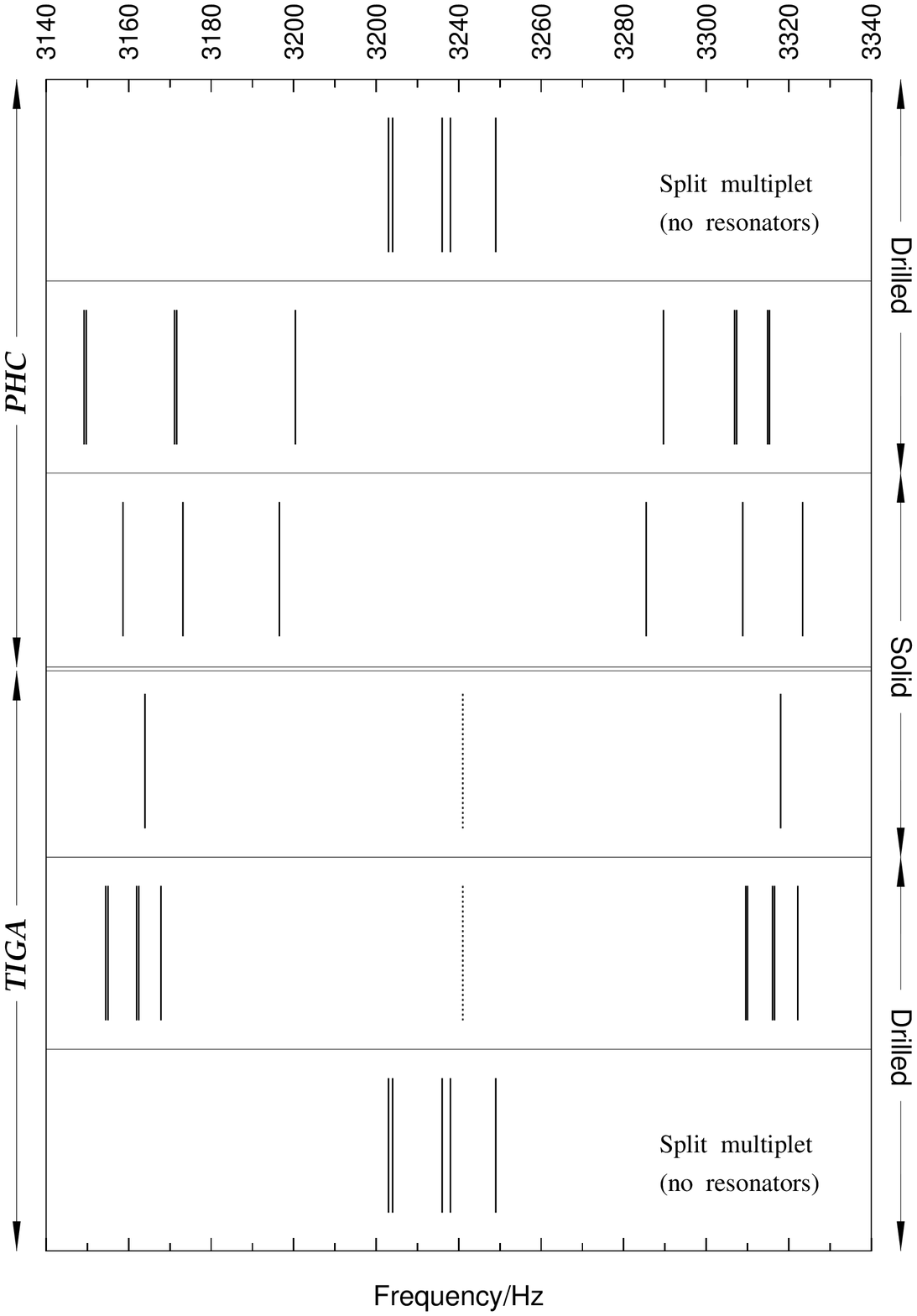,height=22cm,width=15cm,rheight=22cm}
\caption{A comparative display of the line spectra of a {\sl TIGA\/} and
a {\sl PHC\/}, both in a hypothetically perfect solid shape (centre) and
in a more realistic, drilled-for-suspension device (bottom and top,
respectively). {\sl TIGA\/} data are real, see Figure \protect\ref{fig8},
while {\sl PHC\/} data are speculative on the basis of reasonable
assumptions ---see text.
\label{fig9}}
\end{figure}

The results are displayed in Figure \ref{fig9}. It is at once apparent
that the structure of the frequency pairs in a {\it drilled\/} {\sl PHC\/}
is quite similar to that of the perfect {\sl PHC\/}, while the drilled
{\sl TIGA\/} pairs are qualitatively different from those of the ideal
{\sl TIGA\/}. This is not particularly surprising given the symmetries
of both layouts, and is an indication that axially symmetric distributions
may be an interesting alternative to more symmetric distributions in a
real system, as their geometry naturally adapts to the suspension device,
and also require {\it fewer\/} transducers.

\subsection{Other mismatched parameters}

We finally devote a few words to assess the system sensitivity to small
mismatches in resonators' masses, locations and frequencies.

\subsubsection{Resonator mass mismatches}

If the {\it masses\/} are slightly non-equal then we can write

\begin{equation}
  M_a = \eta{\cal M}\,(1+\mu_a\,\eta^{1/2})\ ,\qquad a=1,\ldots,J
  \label{8.9}
\end{equation}
where $\eta\/$ can be defined e.g. as the ratio of the {\it average\/}
resonator mass to the sphere's mass. It is immediately obvious from
equation ({\ref{8.9}) that mass non-uniformities of the resonators only
affect our equations in {\it second order\/}, since resonator mass
non-uniformities result, as we see, in correcctions of order $\eta^{1/2}$
to $\eta^{1/2}$ itself, which is the very parameter of the perturbative
expansions. The system is thus clearly {\it robust\/} to mismatches in
the resonator masses of the type (\ref{8.9}).

\subsubsection{Errors in resonator locations}

The same happens if the {\it locations\/} of the resonators have tolerances
relative to a {\it pre-selected\/} distribution. For let ${\bf n}_a\/$ be a
set of resonator locations, for example the {\sl TIGA\/} or the {\sl PHC\/}
positions, and let ${\bf n'}_a\/$ be the real ones, close to the former:

\begin{equation}
  {\bf n'}_a = {\bf n}_a\,(1+{\bf v}_a\,\eta^{1/2})\ ,\qquad a=1,\ldots,J
  \label{8.10}
\end{equation}

The values ${\bf n}_a$ determine the eigenvalues $\zeta_a\/$ in equation
(\ref{5.2}), and also they appear as arguments to the spherical harmonics
in the system response functions of sections 6 and 7. It follows from
(\ref{8.10}) by continuity arguments that

\begin{mathletters}
\label{8.105}
\begin{eqnarray}
  Y_{lm}({\bf n'}_a) & = & Y_{lm}({\bf n}_a) + O(\eta^{1/2})
  \label{8.105.a} \\
  \zeta'_a & = & \zeta_a + O(\eta^{1/2})
  \label{8.105.b}
\end{eqnarray}
\end{mathletters}

Inspection of the equations of sections 5--7 shows that both $\zeta_a\/$
and $Y_{lm}({\bf n}_a)$ {\it always\/} appear within lowest order terms,
and hence that corrections to them of the type (\ref{8.105}) will affect
those terms in {\it second order\/} again. We thus conclude that the
system is also {\it robust\/} to small misalignments of the resonators
relative to pre-established positions.

\subsubsection{Resonator frequency mistunings}

The resonator {\it frequencies\/} may also differ amongst them, so let

\begin{equation}
  \Omega_a = \Omega\,(1+\rho_a\,\eta^{1/2})\ ,\qquad a=1,\ldots,J
  \label{8.11}
\end{equation}

To assess the consequences of this, however, we must go back to equation
(\ref{3.18}) and see what the coefficients in its series solutions of the
type (\ref{4.11.a}) are. The procedure is very similar to that of section
\ref{s8.1}, and will not be repeated here; the lowest order coefficient
$\chi_\frac{1}{2}$ is seen to satisfy the algebraic equation

\begin{equation}
  \det\left[\delta_{ab} - \frac{1}{\chi_\frac{1}{2}}\,\sum_{c=0}^{J}\,
  \frac{\chi_{ac}^{(n_0l_0)}\,\delta_{cb}}{\chi_\frac{1}{2}-\rho_c}\right]
  = 0	  \label{8.12}
\end{equation}

which reduces to (\ref{5.1}) when all the $\rho\/$'s vanish, as expected.
This appears to potentially have significant effects on our results to
lowest order in $\eta^{1/2}$, but a more careful consideration of the facts
shows that it is probably unrealistic to think of such large tolerances in
resonator manufacturing as implied by equation (\ref{8.11}) in the first
place. In the {\sl TIGA\/} experiment, for example \cite{phd}, an error of
order $\eta^{1/2}$ would amount to around 50 Hz of mistuning between
resonators, an absurd figure by all means. In a full scale sphere
($\sim$40 tons, $\sim$3 metres in diameter, $\sim$800 Hz fundamental
quadrupole frequency, $\eta\/$\,$\sim$\,10$^{-5}$) the same error would
amount to between 5 Hz and 10 Hz in resonator mistunings for the lowest
frequency. This is probably excessive for a capacitive transducer, but may
be realistic for an inductive one. With this exception, it is thus more
appropriate to consider that resonator mistunings are at least of order
$\eta\/$. If this is the case, though, we see once more that the system
is quite insensitive to such mistunings.

Summing up the results of this section, we can say that the resonator
system dynamics is quite {\it robust\/} to small (of order $\eta^{1/2}$)
changes in its various parameters. The important exception is of course
the effect of suspension drillings, which do result in significant changes
relative to the ideally perfect device, but which can be relatively easily
calculated. This theoretical picture is fully supported by experiment, as
{\it robustness\/} in the parameters here considered has been reported in
reference \cite{jm97}.

\section{Conclusions}

A spherical GW antenna is a natural multimode device with very rich
potential capabilities to detect GWs on earth. But such detector is not
just a bare sphere, it requires a set of {\it motion sensors\/} to be
practically useful. It appears that transducers of the {\it resonant\/}
type are the best suited ones for an efficient performance of the detector.
{\it Resonators\/} however significantly {\it interact\/} with the sphere,
and they affect in particular its frequency spectrum and vibration modes
in a specific fashion, which has to be properly understood before reliable
conclusions can be drawn from the system readout.

The main objective of this paper has been the construction and development
of an elaborate theoretical model to describe the joint dynamics of a solid
elastic sphere and a set of {\it radial motion\/} resonators attached to
its surface at arbitrary locations, with the purpose to make predictions
of the system characteristics and response, in principle with arbitrary
mathematical precision.

We have shown that the solutions to our equations of motion must necessarily
be given as an ascending series in powers of the small ``coupling constant''
$\eta\/$, the ratio of the average resonator mass to the mass of the large
sphere. The {\it lowest order\/} approximation coresponds to terms of order
$\eta^{1/2}$ and, to this order, we recover, and widely generalise, other
authors' results \cite{jm97,grg,ts}, obtained by them on the basis of
certain simplifying assumptions. This has in particular enabled us to
assess the system response for arbitrary resonator layouts, and to search
the equations for configurations other than the highly symmetric {\sl TIGA\/},
whose properties and/or performance may thus be comparatively assessed. This
search has led us to make a specific proposal, the {\sl PHC\/}, which is
based on a pentagonally symmetric set of 5 rather than 6 resonators per
quadruopole mode sensed. This {\sl PHC\/} distribution presents a somewhat
wider frequency spectrum than the {\sl TIGA\/}, and has the interesting
property that different {\it wave amplitudes\/} selectively couple to
different {\it detector modes\/} having different frequencies, so that the
antenna's mode channels come at different rather than equal frequencies.
The {\sl PHC\/} philosophy can be extended to make a {\it multifrequency\/}
system by using resonators tuned to the first two quadrupole harmonics of
the sphere {\it and\/} to the first monopole, an altogether 11 transducer
set \cite{ls}.

The assessment of {\it symmetry failure\/} effects, as well as other
parameter departures form ideality, has also interested us here. This is
seen to receive a particularly clear treatment in our general scheme: the
theory transparently shows that the system is {\it robust\/} against
relative disturbances of order $\eta\/$ or smaller in any system parameters,
and provides a systematic procedure to assess larger tolerances ---up to
order $\eta^{1/2}$. The system is shown to still be robust to tolerances of
this order in some of its parameters, whilst it is not to others. Included
in the latter group is the effect of spherical symmetry breaking due to
system suspension in the laboratory, which causes {\it degeneracy lifting\/}
of the sphere's eigenfrequencies, now split up into multiplets. By using
our algorithms we have suceeded in numerically {\it reproducing\/} the
reportedly measured frequencies of the {\sl LSU\/} prototype antenna with
fully satisfactory precision. The experimentally reported robustness of the
system to resonator mislocations \cite{jm97} is also in full agrement with
our theoretical predictions.

We take this numerical success as a strong indication that our model
{\it is\/} correct. Beyond that, though, we also feel it sheds much light
into the {\it principles\/} of the functioning of a spherical GW antenna
with resonant transducers, as every result ultimately follows from very
general and fundamental principles, i.e., equations (\ref{2.6}).
In addition, our master equations (\ref{3.16}) are actually {\it simpler\/}
than e.g. Merkowitz and Johnson's \cite{jm93,jm97}, since they are fewer in
number, yet they contain {\it more\/} information about the system, as they
are accurate to arbitrary order in $\eta\/$. {\it Lowest order\/} solutions
to these equations are thus already simpler, and this has an obvious
{\it practical\/} value. But {\it higher order\/} corrections can, and must,
be estimated by suitable analysis of those equations. We have not attempted
to comprehensively discuss fine structure corrections in this paper, which
will likely be necessary as experimental techniques improve. We defer them
to near future work.

\acknowledgements{
We are indebted with Stephen Merkowitz for his kind supply of the
{\sl TIGA\/} prototype data, without which a significant part of this
work would have been speculative. Fruitful discusions with him are also
gratefully acknowledged. We thank Eugenio Coccia for invaluable interaction
and encouragement throughout the development of this research, and for the
organisation of very interesting seminars in Rome to more openly discuss
the topics of this paper. Thanks are also due to Curt Cutler for pointing
out to us an initial error in equation (\ref{2.1.b}). We have received
financial support from the Spanish Ministry of Education through contract
number PB93-1050.}

\appendix

\section{}

We give in this Appendix a few important properties of the matrix
$P_l({\bf n}_a\!\cdot\!{\bf n}_b)$ for arbitrary $l\/$ and resonator
locations ${\bf n}_a\/$ ($a\/$=1,...,$J\/$) which are useful for detailed
system resonance characterisation.

We first note the {\it summation formula\/} for spherical harmonics
\cite{Ed60}

\begin{equation}
  \sum_{m=-l}^l\,Y_{lm}^*({\bf n}_a)\,Y_{lm}({\bf n}_b) = 
  \frac{2l+1}{4\pi}\,P_l({\bf n}_a\!\cdot\!{\bf n}_b)\ ,\qquad
  a,b=1,\ldots,J        \label{A.1}
\end{equation}

with an obvious change of notation in the arguments to the spherical
harmonics, and where $P_l\/$ is a Legendre polynomial:

\begin{equation}
  P_l(z) = \frac{1}{2^l\,l!}\,\frac{d^l}{dz^l}\,(z^2-1)^l   \label{A.15}
\end{equation}

To ease the notation we shall use the symbol ${\cal P}_l\/$ to mean the
entire $J\/$$\times$$J\/$ matrix $P_l({\bf n}_a\!\cdot\!{\bf n}_b)$, and
introduce Dirac {\it kets\/} $|m\rangle$ for the column $J\/$-vectors

\begin{equation}
  |m\rangle\equiv\sqrt{\frac{4\pi}{2l+1}}\left(\begin{array}{c}
    Y_{lm}({\bf n}_1) \\ \vdots  \\ Y_{lm}({\bf n}_J)
  \end{array}\right)\ \ ,\qquad m=-l,\ldots,l
  \label{A.2}
\end{equation}

These kets are {\it not\/} normalised; in terms of them equation (\ref{A.1})
can be rewritten in the more compact form

\begin{equation}
  {\cal P}_l = \sum_{m=-l}^l\,|m\rangle\langle m|     \label{A.3}
\end{equation}

Equation (\ref{A.3}) indicates that the {\it rank\/} of the matrix
${\cal P}_l\/$ cannot exceed $(2l+1)$, as there are only $(2l+1)$ kets
$|m\rangle$. So, if $J\/$\,$>$\,$(2l+1)$ then it has at least
$(J-2l-1)$ identically null eigenvalues ---there can be more if some of
the ${\bf n}_a\/$'s are parallel, as this causes rows (or columns) of
${\cal P}_l\/$ to be repeated.

We now prove that the non-null eigenvalues of ${\cal P}_l\/$ are
{\it positive\/}. Clearly, a regular eigenvector, $|\phi\rangle$, say, of
${\cal P}_l\/$ will be a linear combination of the kets $|m\rangle$:

\begin{equation}
  {\cal P}_l \,|\phi\rangle = \zeta^2\,|\phi\rangle\ ,\qquad
  |\phi\rangle = \sum_{m=-l}^l\,\phi_m\,|m\rangle
  \label{A.4}
\end{equation}
where we have called $\zeta^2$ the corresponding eigenvalue, since it is a
positive number, as we shall shortly prove. If the second (\ref{A.4}) is
substituted into the first then it is immediately seen that

\begin{equation}
  \sum_{m'=-l}^l\,\left(\zeta^2\,\delta_{mm'} -
  \langle m|m'\rangle\right)\,\phi_{m'} = 0    \label{A.5}
\end{equation}
which admits non-trivial solutions if and only if

\begin{equation}
  \det\,\left(\zeta^2\,\delta_{mm'} -
  \langle m|m'\rangle\right) = 0    \label{A.6}
\end{equation}

In other words, $\zeta^2$ are the eigenvalues of the
$(2l+1)$$\times$$(2l+1)$ matrix $\langle m|m'\rangle$, which is positive
definite because so is the ``scalar product'' $\langle\phi|\phi'\rangle$.
All of them are therefore strictly positive.

Finally, since the {\it trace\/} is an invariant property of a matrix, and

\begin{equation}
  \text{trace}({\cal P}_l) \equiv \sum_{a=1}^J\,
  P_l({\bf n}_a\!\cdot\!{\bf n}_a) = \sum_{a=1}^J\,1 = J
  \label{A.7}
\end{equation}

we see that the eigenvalues $\zeta_a^2$ add up to $J\/$:

\begin{equation}
  \text{trace}({\cal P}_l) =
  \sum_{a=1}^J\,\zeta_a^2\equiv\sum_{\zeta_a\neq 0}\,\zeta_a^2 = J
\end{equation}


\begin{references}
\bibitem{foot1} We use here the better adapted notation $a_{n0}$ and $a_{n2}$
	instead of $a_n\/$ and $b_n\/$, respectively, of reference
	\protect\cite{lobo}. Numerical values are however identical.
\bibitem{foot2} Typically, such formulas contain higher order terms in
	$\eta\/$ which are calculated on the assumption that only tuned
	modes couple to resonators. They are therefore in error by relative
	magnitudes of order $\eta^{1/2}$.
\end{references}
\end{document}